\documentclass[fleqn,usenatbib]{mnras}

\usepackage{newtxtext,newtxmath}

\usepackage[T1]{fontenc}

\DeclareRobustCommand{\VAN}[3]{#2}
\let\VANthebibliography\thebibliography
\def\thebibliography{\DeclareRobustCommand{\VAN}[3]{##3}\VANthebibliography}

\usepackage{graphicx}	
\usepackage{amsmath}	
\usepackage{longtable}
\usepackage{tabularray}
\usepackage{multicol}
\usepackage{supertabular}

\def\orcid#1{\kern .08em\href{https://orcid.org/#1}{\includegraphics[keepaspectratio,width=0.7em]{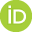}}}

\title[Simultaneous multicolour transit photometry of hot Jupiters]{Simultaneous multicolour transit photometry of hot Jupiters HAT-P-19b, HAT-P-51b, HAT-P-55b, and HAT-P-65b}

\author[Kang et al.]{
H.~Kang$^{1,2}$\orcid{0000-0003-0312-5397},
G.~Chen$^{1,3}$\thanks{E-mail: guochen@pmo.ac.cn}\orcid{0000-0003-0740-5433},
E.~Pall\'e$^{4,5}$\orcid{0000-0003-0987-1593},
F.~Murgas$^{4,5}$\orcid{0000-0001-9087-1245},
N.~Abreu~Garc\'ia$^{4,5}$\orcid{0009-0002-5067-5463},
J.~de~Leon$^{6}$\orcid{0000-0002-6424-3410},
\newauthor
G.~Enoc$^{4,5}$\orcid{0000-0003-0597-7809},
E.~Esparza-Borges$^{4,5}$\orcid{0000-0002-2341-3233},
I.~Fukuda$^{6}$\orcid{0000-0002-9436-2891},
A.~Fukui$^{7,4}$\orcid{0000-0002-4909-5763},
D.~Gal\'an$^{4,5}$\orcid{0000-0001-6191-8251},
Y.~Hayashi$^{6}$\orcid{0000-0001-8877-0242},
\newauthor
K.~Isogai$^{8,6}$\orcid{0000-0002-6480-3799},
T.~Kagetani$^{6}$\orcid{0000-0002-5331-6637},
K.~Kawauchi$^{9,6}$\orcid{0000-0003-1205-5108},
J.~Korth$^{10}$\orcid{0000-0002-0076-6239},
J.~H.~Livingston$^{11,12,13}$\orcid{0000-0002-4881-3620},
R.~Luque$^{14}$\orcid{0000-0002-4671-2957},
\newauthor
Y.~Ma$^{1}$, 
A.~Madrigal-Aguado$^{4}$\orcid{0000-0002-9510-0893},
P.~Meni$^{4,5}$\orcid{0009-0001-7943-0075},
P.~Monta\~nes~Rodriguez$^{15,4}$\orcid{0000-0002-6855-9682},
M.~Mori$^{6}$\orcid{0000-0003-1368-6593},
\newauthor
S.~Mu\~noz~Torres$^{4}$\orcid{0000-0003-4269-4779},
N.~Narita$^{7,11,4}$\orcid{0000-0001-8511-2981},
J.~Orell-Miquel$^{4,5}$\orcid{0000-0003-2066-8959},
H.~Parviainen$^{5,4}$\orcid{0000-0001-5519-1391},
A.~Pel\'aez-Torres$^{4,5}$\orcid{0000-0001-9204-8498},
\newauthor
M.~Stangret$^{16}$\orcid{0000-0002-1812-8024},
M.~Tamura$^{17,11,12}$\orcid{000-0002-6510-0681},
and N.~Watanabe$^{6}$\orcid{0000-0002-7522-8195}
\\
\\
$^{1}$CAS Key Laboratory of planetary sciences, Purple Mountain Observatory, Chinese Academy of Sciences, Nanjing 210023, China\\
$^{2}$School of Astronomy and Space Science, University of Science and Technology of China, Hefei 230026, China\\
$^{3}$CAS Center for Excellence in Comparative Planetology, Hefei 230026, China\\
$^{4}$Instituto de Astrof\'isica de Canarias, V\'ia L\'actea s/n, 38205 La Laguna, Tenerife, Spain\\
$^{5}$Departamento de Astrof\'isica, Universidad de La Laguna, C/ Padre Herrera, 38206 La Laguna, Tenerife, Spain\\
$^{6}$Department of Multi-Disciplinary Sciences, Graduate School of Arts and Sciences, The University of Tokyo, 3-8-1 Komaba, Meguro, Tokyo 153-8902, Japan\\
$^{7}$Komaba Institute for Science, The University of Tokyo, 3-8-1 Komaba, Meguro, Tokyo 153-8902, Japan\\
$^{8}$Okayama Observatory, Kyoto University, 3037-5 Honjo, Kamogatacho, Asakuchi, Okayama 719-0232, Japan\\
$^{9}$Department of Physical Sciences, Ritsumeikan University, Kusatsu, Shiga 525-8577, Japan\\
$^{10}$Lund Observatory, Division of Astrophysics, Department of Physics, Lund University, Box 43, 22100 Lund, Sweden\\
$^{11}$Astrobiology Center, 2-21-1 Osawa, Mitaka, Tokyo 181-8588, Japan\\
$^{12}$National Astronomical Observatory of Japan, 2-21-1 Osawa, Mitaka, Tokyo 181-8588, Japan\\
$^{13}$Astronomical Science Program, Graduate University for Advanced Studies, SOKENDAI, 2-21-1, Osawa, Mitaka, Tokyo 181-8588, Japan\\
$^{14}$Department of Astronomy \& Astrophysics, University of Chicago, Chicago, IL 60637, USA\\
$^{15}$School of Architecture, Universidad Europea de Canarias, C/ Inocencio Garcia, 1 38300, La Orotava, Tenerife, Spain\\
$^{16}$INAF - Osservatorio Astronomico di Padova, Vicolo dell’Osservatorio 5, 35122 Padova, Italy\\
$^{17}$Department of Astronomy, University of Tokyo, 7-3-1 Hongo, Bunkyo-ku, Tokyo 113-0033, Japan
}

\date{Accepted XXX. Received YYY; in original form ZZZ}

\pubyear{2023}

\begin{document}
\label{firstpage}
\pagerange{\pageref{firstpage}--\pageref{lastpage}}
\maketitle

\begin{abstract}
Accurate physical parameters of exoplanet systems are essential for further exploration of planetary internal structure, atmospheres, and formation history. We aim to use simultaneous multicolour transit photometry to improve the estimation of transit parameters, to search for transit timing variations (TTVs), and to establish which of our targets should be prioritised for follow-up transmission spectroscopy. We performed time series photometric observations of 12 transits for the hot Jupiters HAT-P-19b, HAT-P-51b, HAT-P-55b, and HAT-P-65b using the simultaneous four-colour camera MuSCAT2 on the Telescopio Carlos S\'{a}nchez. We collected 56 additional transit light curves from TESS photometry. To derive transit parameters, we modelled the MuSCAT2 light curves with Gaussian processes to account for correlated noise. To derive physical parameters, we performed EXOFASTv2 global fits to the available transit and radial velocity data sets, together with the Gaia DR3 parallax, isochrones, and spectral energy distributions. To assess the potential for atmospheric characterisation, we compared the multicolour transit depths with a flat line and a clear atmosphere model. We consistently refined the transit and physical parameters. We improved the orbital period and ephemeris estimates, and found no evidence for TTVs or orbital decay. The MuSCAT2 broadband transmission spectra of HAT-P-19b and HAT-P-65b are consistent with previously published low-resolution transmission spectra. We also found that, except for HAT-P-65b, the assumption of a planetary atmosphere can improve the fit to the MuSCAT2 data. In particular, we identified HAT-P-55b as a priority target among these four planets for further atmospheric studies using transmission spectroscopy.
\end{abstract}

\begin{keywords}
planets and satellites: individual: (HAT-P-19b, HAT-P-51b, HAT-P-55b, HAT-P-65b) -- planets and satellites: fundamental parameters -- planetary systems -- techniques: photometric
\end{keywords}



\section{Introduction}

\label{sec:introduction}
Since the discovery of the first exoplanet around a Sun-like star \citep{1995Natur.378..355M}, more than 5500 exoplanets have been found, three-quarters of them by transit. When an exoplanet transits its host star, part of the starlight is blocked by the planets in the line of sight of the observer, and the starlight passes through the day-night terminator of the exoplanet's atmosphere. With the flux variation of the planetary system, the transit parameters of the system can be calculated from the transit light curves, which carry information about the internal structure and formation process of the exoplanets \citep{2003ApJ...585.1038S,2007ApJ...659.1661F,2016ApJ...832...41M}. Because the atmospheric opacity varies in different passbands, the properties of the planetary atmosphere can be studied through transmission spectra \citep{2000ApJ...537..916S,2002ApJ...568..377C}, potentially linking the atmospheric chemistry to the planet's formation history and habitability \citep{2014ApJ...794L..12M, 2016SSRv..205..285M, 2016ApJ...832...41M}.

Meaningful investigations of planetary internal structure, atmospheric properties, atmospheric evolution, formation, and migration histories all require precise determination of orbital and physical parameters for the planetary systems as input. In particular, precise transit parameters, together with the latest parallaxes provided by Gaia and additional constraints from spectral energy distributions and stellar evolution models, can improve the estimates of the physical parameters. In the past decade, simultaneous multi-channel imagers, such as GROND \citep{2008PASP..120..405G} and MuSCAT1/2/3 \citep{2015JATIS...1d5001N,2019JATIS...5a5001N,2020SPIE11447E..5KN}, have been extensively used to conduct multicolour follow-up transit photometry. The simultaneous multicolour capability not only allows precise measurements of colour-independent transit parameters to revise physical parameters and to search for transit timing variations (TTVs), but also helps validate candidate planets orbiting faint stars, constrain starspot properties and stellar obliquity, and provide a preliminary assessment of planetary atmospheres \citep[e.g.,][]{2014MNRAS.443.2391M,2014A&A...563A..40C,2019A&A...630A..89P}.

To refine orbital and physical parameters for the known hot Jupiter systems and to prioritise targets for future spectroscopic atmospheric characterisation, we initiated a multicolour transit photometry observing campaign using the four-colour simultaneous camera MuSCAT2 \citep{2019JATIS...5a5001N} in the $g$, $r$, $i$, and $z_s$ bands. In previous studies, we have found evidence for scattering features in the atmospheres of the hot Jupiters WASP-74b \citep{2020A&A...642A..50L} and WASP-104b \citep{2021MNRAS.500.5420C} by combining the MuSCAT2 photometric measurements with those from transit spectrophotometry. Here, we present the MuSCAT2 transit observations for four Saturn-mass hot Jupiters, HAT-P-19b, HAT-P-51b, HAT-P-55b, and HAT-P-65b, complemented by archival data from the Transiting Exoplanet Survey Satellite (TESS). 

HAT-P-19b is a Saturn-mass planet discovered by \citet{2011ApJ...726...52H}, with a mass of 0.29 $M_\mathrm{J}$, a radius of 1.13 $R_\mathrm{J}$, and an equilibrium temperature of 1010~K, orbiting a $V = 12.9$~mag star (0.84 $M_{\odot}$, 0.82 $R_{\odot}$, $T_\mathrm{eff}=4990$~K, and $\mathrm{[Fe/H]} = 0.23$) every 4.009~days. Follow-up transit photometry found no evidence for TTVs \citep{2015MNRAS.451.4060S,2018IBVS.6243....1M,2020MNRAS.496.4174B,2022AJ....164..220H}, except for \citet{2022AJ....164..220H}'s recent study suggesting an orbital decay rate of $-57.7\pm 7.3$~ms\,yr$^{-1}$. The atmosphere of HAT-P-19b has been studied by \citet{2015AA...580A..60M}, who observed one transit with OSIRIS at the Gran Telescopio Canarias (GTC) and derived a flat featureless transmission spectrum. 

HAT-P-51b was discovered by \citet{2015AJ....150..168H}. It is also a Saturn-mass planet, with a mass of 0.31 $M_\mathrm{J}$, a radius of 1.29 $R_\mathrm{J}$, and an equilibrium temperature of 1192~K, orbiting a $V = 13.4$~mag star (0.98 $M_{\odot}$, 1.04 $R_{\odot}$, $T_\mathrm{eff}=5449$~K, and $\mathrm{[Fe/H]}=0.27$) every 4.218~days. 

HAT-P-55b was discovered by \citet{2015PASP..127..851J}, with a mass of 0.58 $M_\mathrm{J}$, a radius of 1.18 $R_\mathrm{J}$, and an equilibrium temperature of 1313~K, orbiting a $V = 13.2$~mag star (1.01 $M_{\odot}$, 1.01 $R_{\odot}$, $T_\mathrm{eff}=5808$~K, and $\mathrm{[Fe/H]}=-0.03$) every 3.585~days. 

HAT-P-65b was discovered by \citet{2016AJ....152..182H}, with a mass of 0.53 $M_\mathrm{J}$, a radius of 1.89 $R_\mathrm{J}$, and an equilibrium temperature of 1930~K, orbiting a $V = 13.1$~mag star (1.21 $M_{\odot}$, 1.86 $R_{\odot}$, $T_\mathrm{eff}=5835$~K, and $\mathrm{[Fe/H]}=0.10$) every 2.605~days. Based on two transits with OSIRIS at the GTC, \citet{2021ApJ...913L..16C} reported the detection of TiO and possible evidence for Na and VO in the atmosphere of HAT-P-65b.

This paper is organised as follows. In Section \ref{sec:data}, we summarise the transit observations and data reduction procedures. In Section \ref{sec:analysis}, we describe the light-curve analysis for MuSCAT2 and TESS, and present the derived transit parameters and orbital period. In Section \ref{sec:physical properties}, we perform the global modelling to refine the physical parameters of the planetary systems. In Section \ref{sec:transmission spectra}, we discuss the wavelength dependence of the transit depth for future atmospheric characterisation. Finally, we draw conclusions in Section \ref{sec:conclusions}.

\section{Observations and data reduction}
\label{sec:data}
\subsection{TCS/MuSCAT2 photometry}
We observed 3 transits of HAT-P-19b, 3 transits of HAT-P-51b, 4 transits of HAT-P-55b, and 2 transits of HAT-P-65b using the four-colour imager MuSCAT2 \citep{2019JATIS...5a5001N} installed on the 1.52 m Telescopio Carlos S\'{a}nchez (TCS) in the Teide Observatory, Tenerife, Spain. MuSCAT2 has the ability to simultaneously observe in four broad passbands: $g$ (400--550 nm), $r$ (550--700 nm), $i$ (700--820 nm), and $z_s$ (820--920 nm). Each passband has its independent CCD and each CCD has 1024$\times$1024 pixels, with a pixel scale of $\sim$0.44 arcsec, giving MuSCAT2 a field of view of $7.4\times7.4$ $\rm arcmin^2$. Due to CCD failure, only three channels were available in some nights. The detailed observation information is given in Table~\ref{tab:obslog}. 

We reduced the MuSCAT2 data with customised IDL scripts as detailed in \citet{2021MNRAS.500.5420C}. In brief, we corrected bias, dark and flat field from the raw images, and performed aperture photometry using the APER routine from DAOPHOT\footnote{\url{https://idlastro.gsfc.nasa.gov/ftp/pro/idlphot/}}. We extracted the central time of each exposure and converted it to Barycentric Julian Dates in Barycentric Dynamical Time \citep[$\mathrm{BJD}_\mathrm{TDB}$;][]{2010PASP..122..935E}. We determined the best aperture radius by minimising the light-curve scatter among a grid of radii ranging from 3 to 32 pixels (equivalent to 1.3$''$--14.1$''$). We also tested different combinations of reference stars to produce the best synthetic reference light curve that minimises the light-curve scatter. The final chosen aperture radii are listed in Table~\ref{tab:obslog}. 

\begin{table*}
\caption[]{Observation summary.}
\begin{center}
\setlength{\tabcolsep}{1.5mm}
\scalebox{1.0}{
\begin{tabular}{cccccccccc}
\hline\hline
\# & Tele. & Instru. & Start night & Start & End & Filter & $t_\mathrm{exp}$ & Airmass$^{\textit{a}}$ & Aperture \\
   &       &         & UT            & UT    & UT  &        & (s)          &             & (pixel)    \\
\hline
\multicolumn{10}{l}{HAT-P-19}\\
1 & TCS & MuSCAT2	&  2018-07-23	& 01:02	 & 05:21	& $ {g, r, i, z_s}$	& 25, 25, 15, 25  & 1.920-1.009-1.009	& 10, 11, 11, 10   \\
2	& TCS	& MuSCAT2	&  2018-08-04	& 00:16	 & 05:33	& $ {g, r, i, z_s}$	& 25, 15, 15, 25  & 1.904-1.007-1.018	& 7, 8, 7, 7   \\
3	& TCS	& MuSCAT2	&  2018-08-08	& 01:01	 & 05:41	& $ {g, r, i, z_s}$	& 10,  6, 10, 15  & 1.444-1.007-1.034	& 13, 14, 11, 11   \\
\hline
\multicolumn{10}{l}{HAT-P-51}\\
1	& TCS	& MuSCAT2	&  2018-10-16	& 22:46	 & 05:19	& $ {g, r, i, z_s}$	& 60,60,15,60 	& 1.119-1.003-1.854	& 15, 12, 11, 11   \\
2	& TCS	& MuSCAT2	&  2018-11-23	& 23:17	 & 03:08	& $ {g, r, i}$		& 30,15,30  	& 1.020-1.020-1.888	& 10, 10, 10   \\
3	& TCS	& MuSCAT2	&  2019-08-04	& 01:04	 & 05:32	& $ {g, r, i}$		& 30,15,30		& 1.918-1.004-1.004	& 11, 10, 11  \\
\hline
\multicolumn{10}{l}{HAT-P-55}\\
1	& TCS	& MuSCAT2	&  2018-05-15	& 23:21	 & 04:04	& $ {g, r, i, z_s}$	& 40-70,20-40,30-60,50-70  & 1.558-1.001-1.025	& 12, 6, 9, 9 \\
2	& TCS	& MuSCAT2	&  2019-06-17	& 21:32	 & 02:54	& $ {r, i, z_s}$	    & 15-18,35,60	  & 1.444-1.001-1.108	& 9, 10, 11 \\
3	& TCS	& MuSCAT2	&  2019-08-17	& 21:00	 & 01:14	& $ {g, r, z_s}$		& 20,15,100 	  & 1.001-1.001-1.792	& 10, 9, 14 \\
4	& TCS	& MuSCAT2	&  2021-06-21	& 22:46	 & 05:05	& $ {g, r, i, z_s}$	& 45,15-30,60,60  & 1.111-1.001-1.880	& 10, 8, 12, 10 \\
\hline
\multicolumn{10}{l}{HAT-P-65}\\
1	& TCS	& MuSCAT2	&  2018-07-29	& 22:03	 & 04:49	& $ {g, r, i, z_s}$	& 30, 30, 30, 40  & 1.654-1.041-1.468	& 11, 11, 11, 11 \\
2	& TCS	& MuSCAT2	&  2019-08-16	& 21:05	 & 01:05	& $ {g, r, z_s}$	    & 40, 20, 120	  & 1.562-1.041-1.053	& 11, 11, 11 \\
\hline
\end{tabular}
}
\begin{flushleft}
    \noindent{\footnotesize{Notes:$^{\textit{a}}$The first and third values refer to the airmass at the beginning and end of the observation. The second value gives the minimum airmass.}}
\end{flushleft}
\label{tab:obslog}  
\end{center}
\end{table*}

\subsection{TESS photometry}
To enlarge the mid-transit time dataset for the refinement of the orbital period and ephemeris, we made use of the archival transit observations conducted by the Transiting Exoplanet Survey Satellite \citep[TESS;][]{2015JATIS...1a4003R}. 
\begin{itemize}
    \item[-] For HAT-P-19b, six transits with an exposure time of 1800 s were observed in full-frame images in Sector 17 between October 8 and November 2, 2019. Another seven transits with an exposure time of 20 s were observed in target pixel files in Sector 57 between September 30 and October 29, 2022. 
    \item[-] For HAT-P-51b, six transits with an exposure time of 1800 s were observed in full-frame images in Sector 17 between October 8 and November 2, 2019. Another six transits with an exposure time of 120 s were observed in full-frame images in Sector 57 between September 30 and October 29, 2022. 
    \item[-] For HAT-P-55b, thirteen transits with an exposure time of 120 s were observed in target pixel files in Sector 25 and 26 between May 14 and July 4, 2020. Another thirteen transits with an exposure time of 120 s were observed in target pixel files in Sector 52 and 53 between May 25 and July 8, 2022, of which two transits were not used in the subsequent analysis due to incomplete transit coverage. 
    \item[-] For HAT-P-65b, ten transits of HAT-P-65b with an exposure time of 120 s were observed in target pixel files in Sector 55 in August 5 and September 1, 2022, and three transits of HAT-P-65b were not used in the subsequent analysis due to bad data quality. 
\end{itemize}

We used the Python package lightkurve \citep{2018ascl.soft12013L} to download the observation data from the MAST data archive\footnote{\url{https://archive.stsci.edu/}}. For HAT-P-19b and HAT-P-51b, the raw light curves in 2019 were created from the tesscut product from full-frame images, and the SPOC light curves \citep{2016SPIE.9913E..3EJ} were used in 2022. For HAT-P-55b, all the raw light curves were created from the target pixel file from tess phot. For HAT-P-65b, the SPOC light curves were used. The adopted time windows were three times the transit duration from the expected transit centre for data with an exposure time of 1800 s, one and a half times for data with 120 s, and one times for data with 20 s. The raw light curves were normalised by the decile value of the whole TESS raw flux of all transits.

\section{Light-curve analysis}
\label{sec:analysis}
We modelled the raw light curves with the transit model from \citet{2002ApJ...580L.171M} using the Python package batman \citep{2015ascl.soft10002K}:
\begin{equation}
\label{eq1}
\mu(t;\theta)=m(t;R_p/R_{\star},T_\mathrm{mid},i,a/R_{\star},u_1,u_2).
\end{equation}
The free parameters of the transit model consist of radius ratio $R_p/R_{\star}$, mid-transit time $T_\mathrm{mid}$, orbital inclination $i$, orbital semi-major axis in units of the stellar radius $a/R_{\star}$, the quadratic limb-darkening coefficients (LDCs) $u_1$ and $u_2$. Circular orbits are adopted for HAT-P-51, HAT-P-55, and HAT-P-65. For HAT-P-19, the orbital eccentricity and argument of periastron were fixed to 0.084 and $256$~deg \citep{2011ApJ...726...52H}, respectively. The orbital period is fixed to literature values in the transit model. For the planetary systems which are diluted by an unresolved companion star, the transit model is revised to $\mu^*(t;\theta,f_\mathrm{c})$ to account for the flux dilution from the companion:
\begin{equation}
\label{eq2}
\mu^*(t;\theta,f_\mathrm{c})=\frac{\mu(t;\theta)+f_\mathrm{c}}{1+f_\mathrm{c}},
\end{equation}
where $\mu(t;\theta)$ is the transit model without the dilution effect, $f_\mathrm{c}$ is the companion-to-target flux ratio.

We employed Gaussian processes (GP) to account for the correlated noise present in the light curves, which was first introduced to transmission spectroscopy by \citet{2012MNRAS.419.2683G}. The one-dimensional GP regression was performed by the Python package celerite \citep{2017AJ....154..220F}, which accepts time series as the input vector. The GP mean function was described by the transit model multiplied by a linear polynomial baseline function $b(\varphi)$. The GP covariance matrix $\boldsymbol{K}$ was described by a combined kernel which consisted of an approximated 3/2-order Matern kernel for time-correlated red noise and a jitter kernel for underestimated white noise:
\begin{equation}
\label{eq3}
\kappa(\tau;\sigma_1,\rho,\sigma_2)=\sigma_1^2\left(1+\frac{\sqrt{3}\tau}{\rho}\right)\exp\left(-\frac{\sqrt{3}\tau}{\rho}\right)+\sigma_2^2\delta,
\end{equation}
where $\tau=|t_i-t_j|$ is the distance between two data points in time, $\rho$ and $\sigma_1^2$ are the length and variance scales of systematic noise, $\sigma_2^2$ is the variance of underestimated white noise.

We performed the affine invariant Markov Chain Monte Carlo (MCMC) ensemble sampler using the Python package emcee \citep{2013ascl.soft03002F} to explore the posterior probability distributions of the free parameters. We adopted uniform priors for most of the transit parameters, polynomial coefficients for the baseline function, and log-uniform priors for the GP hyperparameters. We imposed normal priors on the LDCs $u_1$ and $u_2$ (see Table~\ref{tab: ldc priors}). We calculated LDCs from the ATLAS stellar atmosphere models by interpolating in the model grids using the stellar parameters \citep{2015MNRAS.450.1879E}. We ran two short chains for the burn-in phase and one long to ensure convergence chain for formal production. 

\begin{figure*}
\centering
\includegraphics[width=0.95\textwidth]{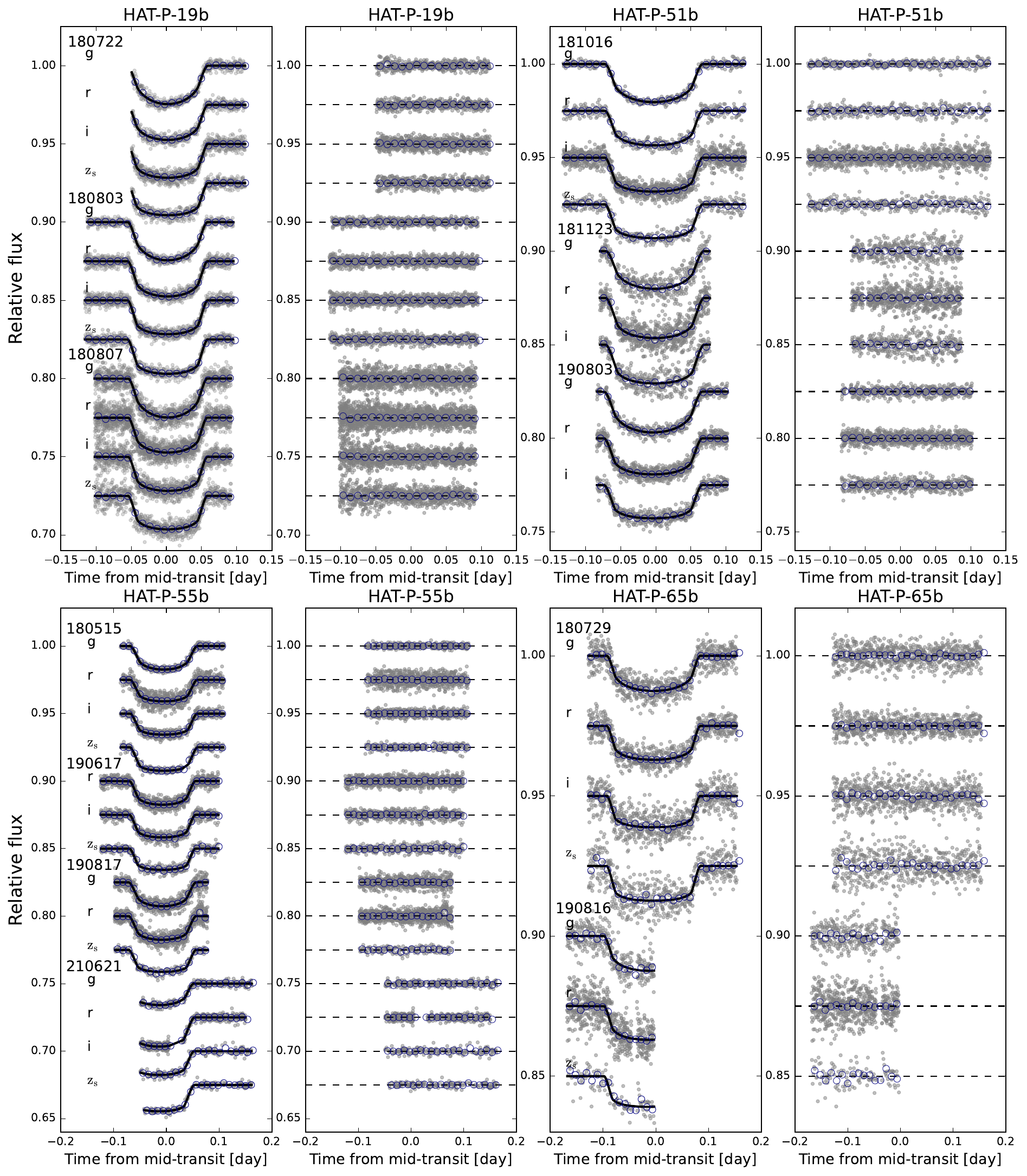}
\caption{MuSCAT2 multicolour transit light curves of HAT-P-19b, HAT-P-51b, HAT-P-55b, and HAT-P-65b. The first and third panels show the light curves after removal of the systematics. The second and fourth panels show the best-fit residuals. The black solid lines show the best-fit model, and the navy circles show the 15-min binned points.}
\label{fig:detrended light curves}
\end{figure*}

\begin{table*}
\renewcommand\arraystretch{1.4}
\centering
\caption{Derived transit parameters and orbital ephemeris.}
\label{tab: white-colour light curves parameters of weighted}
\scalebox{0.95}{
\begin{tabular}{ccccccc}
\hline\hline
Planet	& Source	& $R_p/R_{\star}$		& $i$ (degree)		& $a/R_{\star}$	& $P$ (days)	& $T_0 ({\rm BJD_{TDB}})$\\
\hline
HAT-P-19b 	&	&	&	&	&	& \\
 & This work	& $0.1346\pm0.0004$	& $89.52\pm0.20$	& $12.84\pm0.06$	& $4.00878322_{-0.00000019}^{+0.00000019}$	& $2456610.863945_{-0.000071}^{+0.000071}$\\
 & \citet{2011ApJ...726...52H}	& $0.1418\pm0.0020$	& $88.2\pm0.4$	 & $12.24\pm0.67$	& $4.008778\pm0.000006$	  & $2455091.53417\pm0.00034 ^{\textit{a}}$\\
 & \citet{2015MNRAS.451.4060S}	& $0.1378\pm0.0014$	& $88.51\pm0.22$ & $12.36\pm0.09$	& $4.0087842\pm0.0000007$	& $2455091.53500\pm0.00015$\\
 & \citet{2015AA...580A..60M}	& $0.1390\pm0.0012$	& $88.89\pm0.32$	& $12.37\pm0.21$	& \text{-}	& \text{-}\\
 & \citet{2018IBVS.6243....1M}   & \text{-}	& \text{-}	& \text{-}	& $4.00878332\pm0.00000059$	& $2455091.53501\pm0.00015$\\
 & \citet{2020MNRAS.496.4174B}	& \text{-}	& $89.11_{-0.29}^{+0.42}$	& $12.66_{-0.20}^{+0.43}$	& $4.00878330\pm0.00000033$	& $2456827.337856\pm0.000085$\\
 & \citet{2022ApJS..259...62I} & \text{-}	& \text{-}	& \text{-}	& $4.00878403\pm0.00000049$	& $2456935.57551\pm0.00012$\\
 & \citet{2022ApJS..258...40K} & \text{-}	& \text{-}	& \text{-}	& $4.0087842\pm0.0000004$	& $2456899.49658\pm0.00010$\\
\hline
HAT-P-51b 	&	&	&	&	&	& \\
 &This work	& $0.1253\pm0.0006$	& $89.28\pm0.30$	& $10.86\pm0.08$	& $4.21802091_{-0.00000066}^{+0.00000066}$	& $2458349.53123_{-0.00014}^{+0.00013}$\\
 & \citet{2015AJ....150..168H}	& $0.1278\pm0.0020$	& $88.48\pm0.57$	& $10.48_{-0.40}^{+0.28}$	& $4.2180278\pm0.0000059$	& $2456194.12204\pm0.00040 ^{\textit{a}}$\\
 & \citet{2022ApJS..258...40K} & \text{-}	& \text{-}	& \text{-}	 & $4.2180226\pm0.0000009$	& $2457868.67797\pm0.00024$\\
\hline
HAT-P-55b 	&	&	&	&	&	&\\
 &This work	&$0.1220\pm0.0007$	&$86.80\pm0.22$	&$9.06\pm0.16$	& $3.58523130_{-0.00000051}^{+0.00000051}$	& $2458311.92255_{-0.00013}^{+0.00013}$\\
 & \citet{2015PASP..127..851J}	& $0.1202\pm0.0019$	& $87.70\pm0.56$	& $9.79\pm0.34$	& $3.5852467\pm0.0000064$	& $2456730.83468\pm0.00027 ^{\textit{a}}$\\
 & \citet{2022ApJS..259...62I} & \text{-}	& \text{-}	& \text{-}	& $3.5852316\pm0.0000010$	& $2458989.53140\pm0.00039$\\
 & \citet{2022ApJS..258...40K} & \text{-}	& \text{-}	& \text{-}	& $3.5852329\pm0.0000012$	& $2457720.3595\pm0.0003$\\
\hline
HAT-P-65b 	&	&	&	&	&	&\\
 & This work	& $0.1006\pm0.0009$	& $88.3\pm1.0$	& $5.18\pm0.07$	& $2.60544751_{-0.00000049}^{+0.00000050}$	& $2458139.35067_{-0.00021}^{+0.00021}$\\	 
 & \citet{2016AJ....152..182H}	& $0.1045\pm0.0024$	& $84.2\pm1.3$	& $4.57\pm0.20$	& $2.6054552\pm0.0000031$	& $2456409.33263\pm0.00046 ^{\textit{a}}$\\
 & \citet{2021ApJ...913L..16C}	& $0.0994\pm0.0025$	& $89.10_{-0.83}^{+0.63}$	& $5.22_{-0.04}^{+0.03}$	& \text{-}	&  \text{-} \\
 & \citet{2022ApJS..258...40K} & \text{-}	& \text{-}	& \text{-}	& $2.6054485\pm0.0000009$	& $2457149.2808\pm0.0004$\\
\hline
\end{tabular}
}
\begin{flushleft}
    \noindent{\footnotesize{Notes:$^{\textit{a}}$Mid-transit times in $\mathrm{BJD}_\mathrm{UTC}$.}}
\end{flushleft}
\end{table*}

\begin{table*}
\renewcommand\arraystretch{1.5}
\centering
\caption{Chromatic radius ratios of four hot Jupiter systems.}
\label{tab: transmission spectra}
\scalebox{1.0}{
\begin{tabular}{ccccc}
\hline\hline
Band (nm)	& \multicolumn{4}{c}{$R_p/R_{\star}$}\\
     	& HAT-P-19b 	&HAT-P-51b 	&HAT-P-55b 	& HAT-P-65b\\
\hline
$g$ (400-550)	& $0.1352\pm0.0009$	& $0.1257\pm0.0011$	& $0.1224\pm0.0011$	&$0.0991\pm0.0016$\\
$r$ (550-700)	& $0.1335\pm0.0007$	& $0.1260\pm0.0010$	& $0.1237\pm0.0008$	&$0.1010\pm0.0013$\\
$i$ (700-820)	& $0.1343\pm0.0007$	& $0.1251\pm0.0009$	& $0.1214\pm0.0010$	&$0.0984\pm0.0021$\\
$z_s$ (820-920)	& $0.1350\pm0.0007$	& $0.1239\pm0.0014$	& $0.1201\pm0.0013$	&$0.1041\pm0.0031$\\
\hline
\end{tabular}
}
\end{table*} 

\subsection{MuSCAT2 light curves}
For the MuSCAT2 light curves, we adopted the baseline function $b(\varphi)$ in the form of:
\begin{equation}
\label{eq4}
b(\varphi)=c_0+c_1x+c_2y+c_3s,
\end{equation}
where $x$ and $y$ are the coordinates of the target, $s$ 
is the full width at half maximum of the target's point spread function. The mean function of GP was $\mu^*(t;\theta,f_\mathrm{c})b(\varphi)$ for HAT-P-65 and $\mu(t;\theta)b(\varphi)$ for the others. HAT-P-65 has a background star located at $3.6''$ in the west according to \citet{2016AJ....152..182H}. The background star could not be spatially resolved by the defocused MuSCAT2 observations. Therefore, we estimated the companion-to-target flux ratios $f_\mathrm{c}$ within the MuSCAT2 passbands, which were 0.0086, 0.0094, 0.0098, 0.0101 for $g$, $r$, $i$, and $z_s$, respectively, based on the GTC OSIRIS measurements presented in \citet{2021ApJ...913L..16C}.

We performed two runs of light-curve modelling for each target. In the first run, we aimed to derive the common transit parameters. We jointly fitted multicolour light curves on a nightly basis. Each night had the same values of $i$, $a/R_{\star}$, and $T_\mathrm{mid}$ for all light curves, and each light curve had independent values of $R_p/R_{\star}$, $u_1$, and $u_2$. The coefficients of the baseline function and the GP hyperparameters were always light-curve-dependent. We reported the weighted mean of $i$, $a/R_{\star}$, and $R_p/R_{\star}$ of all nights as the final updated values in Table~\ref{tab: white-colour light curves parameters of weighted}. The detrended MuSCAT2 light curves and best-fit residuals are shown in Fig.~\ref{fig:detrended light curves}.

In the second run, we attempted to evaluate the potential variation in transit depth as a function of wavelength. We fitted each light curve individually and fixed the values of $i$, $a/R_{\star}$, and $T_\mathrm{mid}$ to those obtained in the first run. The free parameters were $R_p/R_{\star}$, $u_1$, $u_2$, baseline function coefficients, and GP hyperparameters. For each passband, the weighted mean of $R_p/R_{\star}$ was taken as the final value and listed in Table~\ref{tab: transmission spectra}.

\subsection{TESS light curves}
For the TESS light curves, we adopted the baseline function $b(\varphi)$ in the form of:
\begin{equation}
\label{eq5}
b(\varphi)=c_0+c_1x+c_2y,
\end{equation}
where $x$ and $y$ are the coordinates of the target. The diluted transit model $\mu^*(t;\theta,f_\mathrm{c})$ was adopted to account for the potential dilution from nearby stars given the large pixel size of TESS. Therefore, $\mu^*(t;\theta,f_\mathrm{mid})b(\varphi)$ was adopted as the GP mean function for all the targets. For HAT-P-19b and HAT-P-51b, we used the supersampling feature of batman to account for the long cadence smearing effect \citep{2010MNRAS.408.1758K}. Since our purpose of modelling the TESS light curves was to measure the mid-transit times, we fixed the radius ratio $R_p/R_{\star}$, the inclination $i$, and the semi-major axis $a/R_{\star}$ to the values obtained from the analysis of the MuSCAT2 light curves. We also fixed the limb darkening coefficients to the pre-calculated values derived from the code of \citet{2015MNRAS.450.1879E}. Each light curve had an independent mid-transit time. The detrended TESS light curves and their best-fit residuals are shown in Fig.~\ref{fig: TESS light curves}. 

\subsection{Transit parameter refinement}
Based on MuSCAT2's light curve analysis of all nights and four passbands in the first run, we are able to refine the transit parameters for the four hot Jupiter systems, which are shown in Table~\ref{tab: white-colour light curves parameters of weighted} along with literature values for comparison. 

For HAT-P-19b, HAT-P-51b, and HAT-P-55b, the transit parameters are derived from at least three MuSCAT2 transits, resulting in smaller uncertainties than those reported in the literature. In the case of HAT-P-65b, only two transits were observed by MuSCAT2, and only one of them covered the entire transit event. The uncertainties of $i$ and $a/R_\star$ are slightly larger than those derived from two GTC transits \citep{2021ApJ...913L..16C}, but still consistent with the latter. 

However, the transit parameters measured in different studies are not exactly in agreement. This discrepancy is likely due to the degeneracy between $i$ and $a/R_\star$, since the measurements from different studies show a correlation trend (i.e., larger $i$ with larger $a/R_\star$) consistent with the $i$-$a/R_\star$ degeneracy. Thanks to the multiple observations and the wide wavelength coverage of MuSCAT2, the transit parameters can be tightly constrained, with colour-dependent bias being eliminated.  

\subsection{Orbital period determination}
We derived the mid-transit time of each transit for all the MuSCAT2 and TESS observations. To investigate the transit timing variations and to improve the orbital ephemeris, we also collected other mid-transit times published in the literature. All the mid-transit times have been converted to the $\rm BJD_{TDB}$ standard and presented in Table~\ref{tab: mid-transit time}. 
\begin{itemize}
    \item For those with raw light curves available \citep{2011ApJ...726...52H, 2015AJ....150..168H, 2016AJ....152..182H, 2021ApJ...913L..16C}, we recalculated their mid-transit times using our light curve analysis method.
    \item For HAT-P-19b, we did not include the mid-transit times in \citet{2020MNRAS.496.4174B}, which have very small error bars and show a general downward offset from the linear ephemeris derived from the other times.
    \item For HAT-P-51b, we discarded the mid-transit time of the 2018-11-23 transit because the computer time of that night was not properly synchronised with the Network Time Protocol server.
    \item For HAT-P-55b and HAT-P-65b, we discarded the the partial transits on 2021-06-21 and 2019-08-16, respectively.
\end{itemize}

The mid-transit times $T_\mathrm{mid}$ were fitted as a function of the epoch $E$ using a linear model and a quadratic model, respectively. For the linear model, the planet was assumed to have a constant orbital period $P$:
\begin{equation}
\label{eq6}
T_\mathrm{mid} = T_0 + PE, 
\end{equation}
where $T_0$ is the mid-transit time at zero epoch. The zero epoch was optimised to give the smallest error bar for $T_0$. For the quadratic model, the planet was assumed to have a decaying orbital period:
\begin{equation}
\label{eq7}
T_\mathrm{mid} = T_0 + PE + \frac{1}{2}\frac{\mathrm{d} P}{\mathrm{d} E}E^2,
\end{equation}
where ${\mathrm{d} P}/{\mathrm{d} E}$ is the decay rate between successive transits. We used the Bayesian Information Criterion ($\mathrm{BIC}=\chi^2 + k\log N$) to perform model comparison, where $k$ is the number of free parameters and $N$ is the number of data points.

\begin{figure*}
\centering
\includegraphics[width=\textwidth]{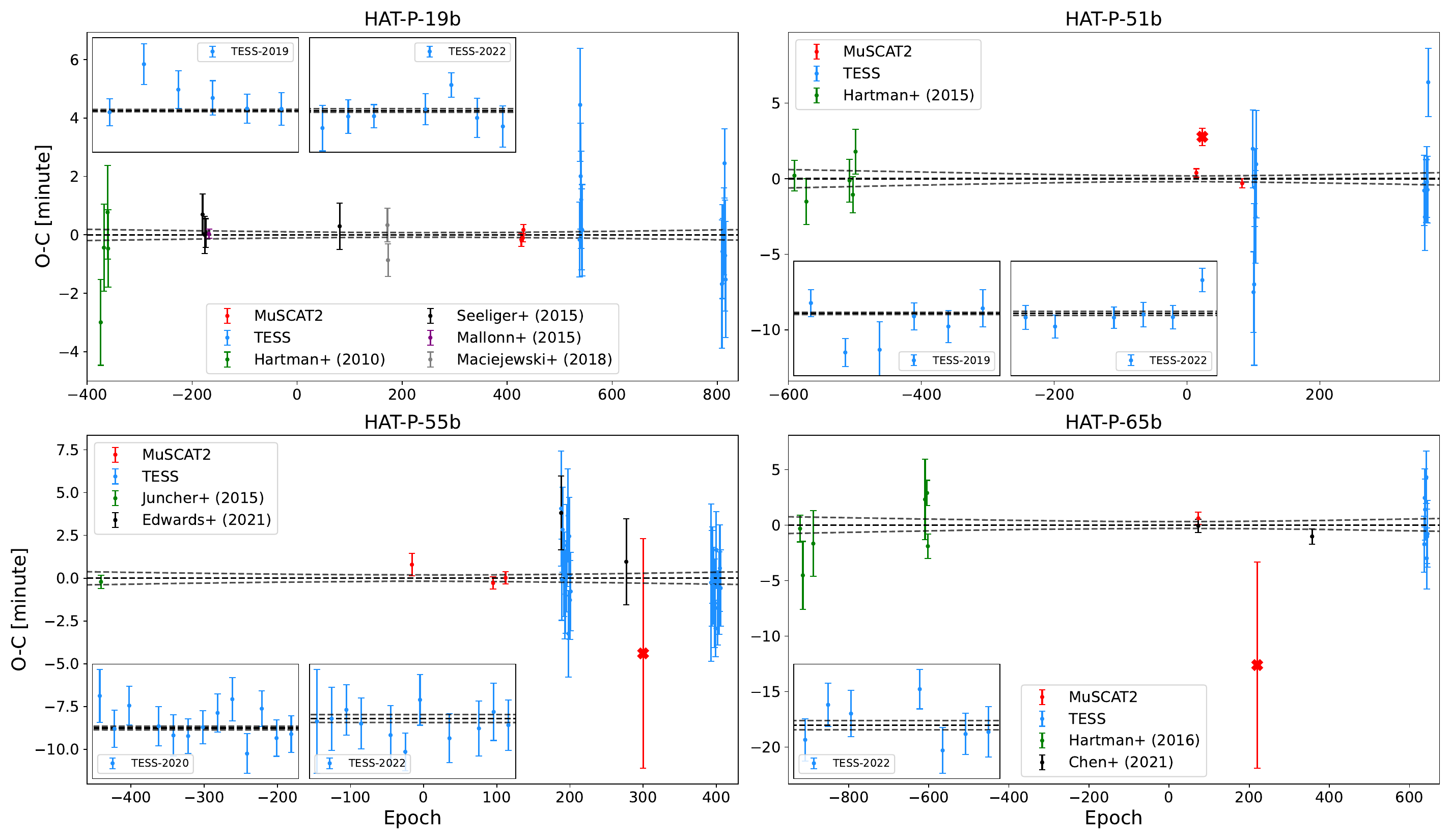}
\caption{Timing residuals of four systems with the constant period model. Each data point is the difference between the observed mid-transit time and the best-fit linear model. The middle dashed line is the zero line, the other two dashed lines show the range of $1\sigma$ uncertainty. The inset shows the zoomed view of the residuals of the TESS light curves. The crossed points were discarded in the period modelling. }
\label{fig: ttv}
\end{figure*}

The results of the model comparison are shown in Table~\ref{tab: model comparing}. For the four planetary systems, the difference between the constant period model and the orbital decay model $\Delta\mathrm{BIC}$ is $-$3.26, $-$2.64, $-$2.34, and $-$2.11, respectively. The constant period model is favored with a lower BIC value in all four planetary systems, indicating that there is no evidence for orbital decay. Therefore, our results do not support the claims of potential orbital decay in HAT-P-19b \citep{2022AJ....164..220H} and HAT-P-51b \citep{2024NewA..10602130Y}. Meanwhile, the timing residuals of the four systems with the best-fit period model show no sign of transit timing variation (Fig.~\ref{fig: ttv}). The refined period and reference ephemeris are given in Table~\ref{tab: white-colour light curves parameters of weighted}.

\begin{table}
\renewcommand\arraystretch{1.5}
\centering
\caption{Comparison of the constant period model and the orbital decay model.}
\label{tab: model comparing}
\scalebox{0.85}{
\begin{tabular}{cccc}
\hline\hline
Parameter	& Symbol	& Constant period  & Orbital decay\\
\hline
HAT-P-19b \\
Number of data  & n & 27 & 27\\
Degrees of freedom  & $dof$  & 24  & 23\\
Chi-square    & $\chi^2$  & 23.48  & 23.45\\
Reduce chi-square     & $\chi^2/dof$  & 0.98  & 1.02\\
Bayesian information criterion  & BIC   & 33.37   & 36.63\\
RMS of residual (second)    &  RMS & 80.8      & 81.1\\
\hline
HAT-P-51b \\
Number of data  & n & 19 & 19\\
Degrees of freedom  & $dof$  & 16  & 15\\
Chi-square statistic    & $\chi^2$  & 27.06  & 26.76\\
Reduce chi-square statistic     & $\chi^2/dof$  & 1.69  & 1.78\\
Bayesian information criterion  & BIC   & 35.89   & 38.53\\
RMS of residual (second)     &  RMS  & 175.5      & 175.7\\
\hline
HAT-P-55b \\
Number of data  & n & 30 & 30\\
Degrees of freedom  & $dof$  & 27  & 26\\
Chi-square statistic    & $\chi^2$  & 19.46 & 18.40\\
Reduce chi-square statistic     & $\chi^2/dof$  & 0.72  & 0.71\\
Bayesian information criterion  & BIC   & 29.66  & 32.00\\
RMS of residual (second)     &  RMS  &  107.2     & 106.3\\
\hline
HAT-P-65b\\
Number of data  & n & 16 & 16\\
Degrees of freedom  & $dof$  & 13  & 12\\
Chi-square statistic    & $\chi^2$  & 22.21 & 21.53\\
Reduce chi-square statistic     & $\chi^2/dof$  & 1.71  & 1.79\\
Bayesian information criterion  & BIC   & 30.52  & 32.63\\
RMS of residual (second)     &  RMS  &  135.9     & 139.6\\
\hline
\end{tabular}
}
\end{table}

\section{Physical properties}
\label{sec:physical properties}
Except for HAT-P-19, no follow-up physical property determinations have been made for these planetary systems. To refine their physical parameters, we used the IDL package EXOFASTv2 \citep{2019arXiv190709480E} to perform a global modelling of the MuSCAT2 transit light curves, the radial velocity (RV) measurements from the literature, the isochrones from the MESA Isochrones and Stellar Tracks \citep[MIST;][]{2016ApJS..222....8D}, and the spectral energy distribution (SED) from broadband photometry. The use of the MIST stellar evolutionary models produces consistent models for both isochrones and SED. The latest stellar parallax from the Gaia third data release \citep[DR3;][]{2022yCat.1355....0G} provides an accurate prior on the stellar distance, which places a tight constraint on the stellar radius in the SED model. The collected broadband photometry and Gaia DR3 stellar parallax are listed in Table \ref{tab: brightness of four hot Jupiters}.

\begin{figure*}
\centering
\includegraphics[width=\textwidth]{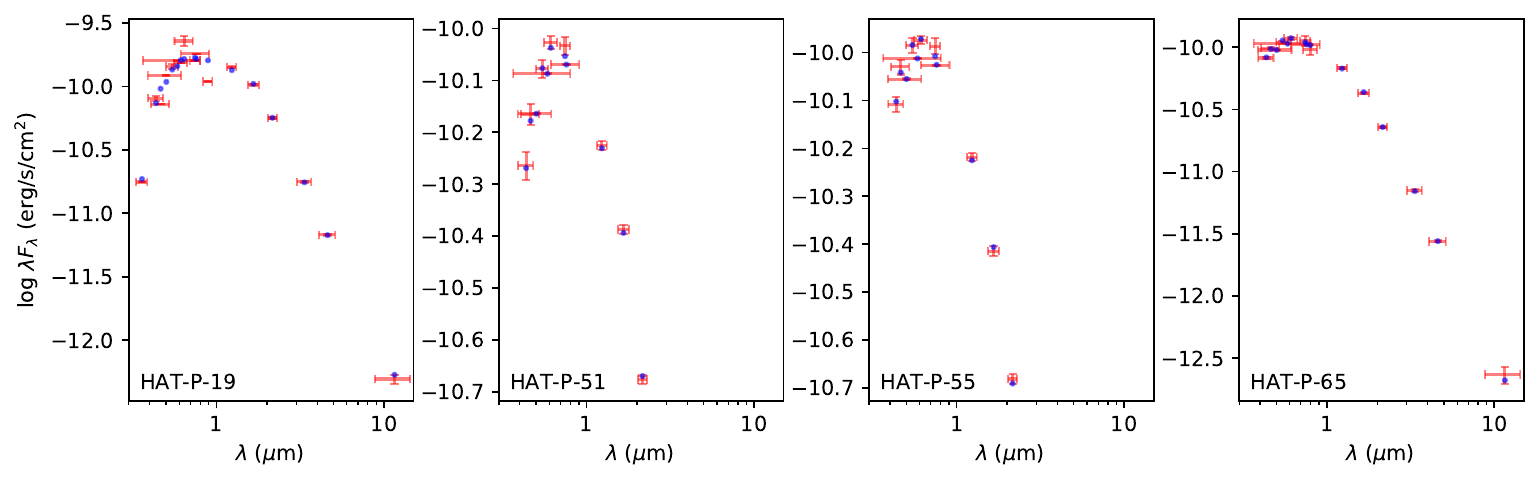}
\caption{Spectral energy distributions (SED) of HAT-P-19, HAT-P-51, HAT-P-55, and HAT-P-65 from broadband photometry. The red data points with error bars are the broadband photometric measurements. The blue circles are the best-fit SED model values. }
\label{fig: sed model}
\end{figure*}

\begin{figure*}
\centering
\includegraphics[width=\textwidth]{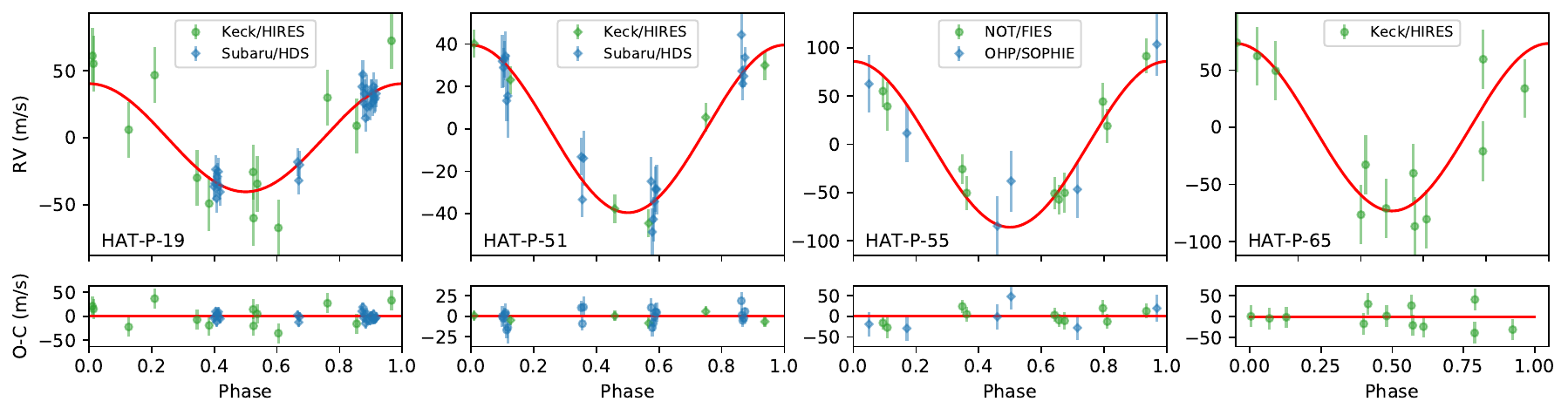}
\caption{Radial velocity (RV) observations of HAT-P-19, HAT-P-51, HAT-P-55, and HAT-P-65. The red curves show the best-fit model from the EXOFASTv2 fit.}
\label{fig: rv model}
\end{figure*}

\begin{figure*}
\centering
\includegraphics[width=\textwidth]{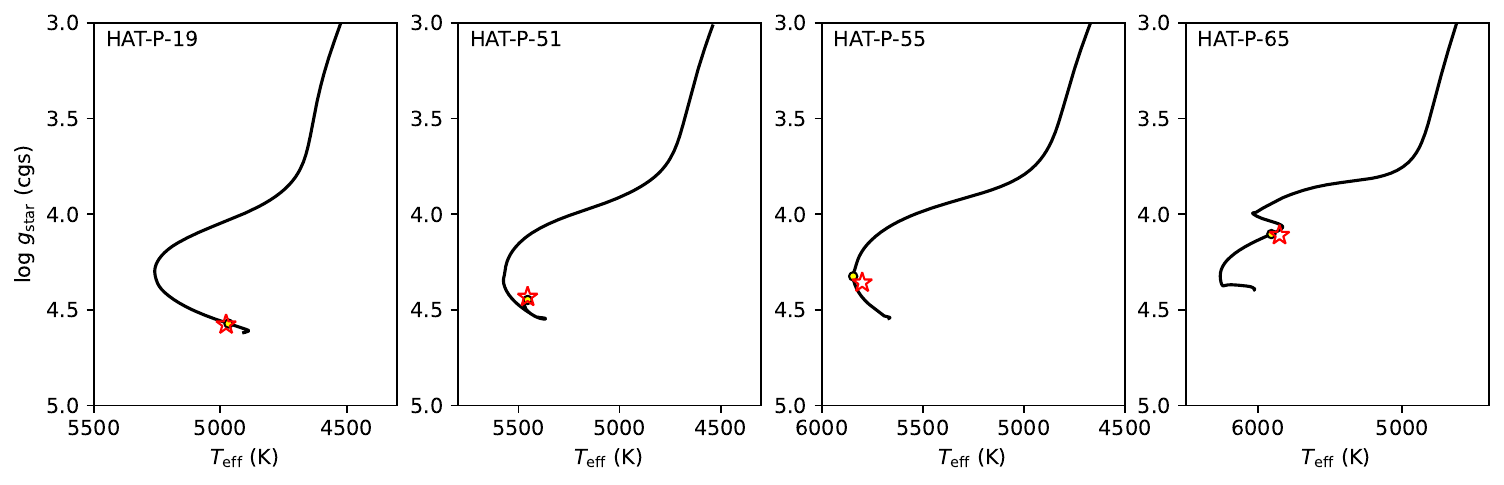}
\caption{Stellar evolutionary tracks of HAT-P-19, HAT-P-51, HAT-P-55, and HAT-P-65. The black line is the MIST mass track interpolated with the best-fit free parameters. The yellow circle highlights the mass track grid closest to the best-fit Equivalent Evolutionary Phase (EEP) value. The red star indicates the best-fit values of $T_{\rm eff}$ and $\log g_{\star}$. }
\label{fig: mist model}
\end{figure*}

We imposed Gaussian priors on the effective temperature $T_{\rm eff}$, the metallicity $\mathrm{[Fe/H]}$, the parallax from Gaia DR3, the quadratic LDCs, the transit period, and placed an upper limit of $3.1E(B-V)_\mathrm{S \& F}$\footnote{https://irsa.ipac.caltech.edu/} on $A_V$. We obtained priors of $T_{\rm eff}$ and $\mathrm{[Fe/H]}$ from  \citet{2011ApJ...726...52H}, \citet{2015AJ....150..168H}, \citet{2015PASP..127..851J}, and \citet{2016AJ....152..182H} and adopted upper limits of $A_V<0.27621$, 0.14849, 0.17019, and 0.27032 for HAT-P-19, HAT-P-51, HAT-P-55, and HAT-P-65, respectively. We ran the MCMC function of EXOFASTv2 to explore the posterior distributions of free parameters and adopted its default convergence criteria (Gelman-Rubin statistics $R_z<1.01$ and independent draws $T_z>1000$). The median and 1-$\sigma$ uncertainties of the derived physical parameters are listed in  Table~\ref{tab: median value}. The best-fit SED models, RV models, and stellar evolutionary models are presented in Fig.~\ref{fig: sed model}, Fig.~\ref{fig: rv model}, and Fig.~\ref{fig: mist model}, respectively. 

For HAT-P-19, we jointly fitted 12 MuSCAT2 light curves along with RV from \citet{2011ApJ...726...52H}. The physical parameters of the HAT-P-19 system have been recently updated by \citet{2020MNRAS.496.4174B}. We derived consistent but marginally smaller values  ($0.277_{-0.016}^{+0.017}$~$M_\mathrm{J}$, $1.008_{-0.013}^{+0.014}$~$R_\mathrm{J}$) for both planetary radius and mass than theirs ($0.284_{-0.017}^{+0.017}$~$M_\mathrm{J}$, $1.064_{-0.034}^{+0.031}$~$R_\mathrm{J}$). 

For HAT-P-51, we jointly fitted 10 MuSCAT2 light curves along with RV from \citet{2015AJ....150..168H}. Our derived planetary mass and radius ($0.307_{-0.020}^{+0.021}$~$M_\mathrm{J}$, $1.205_{-0.016}^{+0.017}$~$R_\mathrm{J}$) are both marginally smaller than those ($0.309_{-0.018}^{+0.018}$~$M_\mathrm{J}$, $1.293_{-0.054}^{+0.054}$~$R_\mathrm{J}$) in the discovery paper \citep{2015AJ....150..168H}.

For HAT-P-55, we jointly fitted 15 MuSCAT2 light curves along with RV from \citet{2015PASP..127..851J}. We derived marginally larger mass and 2.4$\sigma$ larger radius ($0.596_{-0.072}^{+0.073}$~$M_\mathrm{J}$, $1.324_{-0.022}^{+0.023}$~$R_\mathrm{J}$) for the planet, compared to those ($0.582_{-0.056}^{+0.056}$~$M_\mathrm{J}$, $1.182_{-0.055}^{+0.055}$~$R_\mathrm{J}$) in the discovery paper \citep{2015PASP..127..851J}. Our larger planetary radius is the result of both the larger radius ratio and the larger stellar radius ($1.105^{+0.018}_{-0.017}$ v.s. $1.011^{+0.036}_{-0.036}$~$R_\odot$).

For HAT-P-65, we jointly fitted 7 MuSCAT2 light curves, two GTC/OSIRIS light curves \citep{2021ApJ...913L..16C}, along with RV from \citet{2016AJ....152..182H}. Our derived planetary mass and radius are marginally larger and 2.1$\sigma$ smaller ($0.554_{-0.091}^{+0.092}$~$M_\mathrm{J}$, $1.611_{-0.024}^{+0.024}$~$R_\mathrm{J}$) than those in the discovery paper ($0.527_{-0.083}^{+0.083}$~$M_\mathrm{J}$, $1.89_{-0.13}^{+0.13}$~$R_\mathrm{J}$). Our smaller planetary radius is the result of both the smaller radius ratio and the smaller stellar radius ($1.666^{+0.024}_{-0.024}$ v.s. $1.860^{+0.096}_{-0.096}$~$R_\odot$). The difference in stellar radius comes from the transit-constrained stellar density, which could be biased by the degeneracy between $i$ and $a/R_\star$, and was not constrained by the partial transits of \citet{2016AJ....152..182H}.

\begin{table*}
\renewcommand\arraystretch{1.7}
\centering
\caption{Stellar and planetary parameters derived from the EXOFASTv2 global fits.}
\label{tab: median value}
\scalebox{1.0}{
\begin{tabular}{cccccc}
\hline\hline
Symbol	& Parameter (Unit)	& HAT-P-19	& HAT-P-51	& HAT-P-55	& HAT-P-65 \\
\hline
Stellar Parameters	&	&\\
$M_{\star}$		& Mass ($M_{\sun}$)	&$0.807^{+0.034}_{-0.030}$	&$0.961^{+0.040}_{-0.036}$	&$1.028^{+0.050}_{-0.048}$	&$1.297^{+0.056}_{-0.053}$\\
$R_{\star}$		& Radius ($R_{\sun}$) &$0.773^{+0.011}_{-0.010}$	  &$0.995^{+0.013}_{-0.013}$   &$1.105^{+0.018}_{-0.017}$			&$1.666^{+0.024}_{-0.024}$\\
$L_{\star}$		& Luminosity ($L_{\sun}$)	&$0.327^{+0.012}_{-0.013}$	&$0.790^{+0.021}_{-0.022}$  &$1.249^{+0.049}_{-0.045}$	&$2.970^{+0.120}_{-0.120}$\\
$\rho_{\star}$	& Density (cgs)			&$2.468^{+0.018}_{-0.033}$	&$1.380^{+0.012}_{-0.020}$	&$1.073^{+0.036}_{-0.034}$	&$0.397^{+0.003}_{-0.006}$\\
$\log{g}$		& Surface gravity (cgs)	&$4.5687^{+0.0066}_{-0.0067}$ &$4.4256^{+0.0069}_{-0.0073}$	&$4.3630^{+0.0140}_{-0.0130}$		&$4.1079^{+0.0068}_{-0.0074}$\\
$T_{\rm eff}$	& Effective Temperature (K)		&$4962^{+39}_{-41}$	  &$5453^{+31}_{-32}$			&$5804^{+37}_{-37}$			&$5872^{+40}_{-40}$\\
$[{\rm Fe/H}]$	& Metallicity (dex)		&$0.166^{+0.070}_{-0.068}$	&$0.302^{+0.052}_{-0.059}$	&$0.003^{+0.049}_{-0.032}$	&$0.208^{+0.050}_{-0.055}$\\
$[{\rm Fe/H}]_{0}$	& Initial Metallicity &$0.156^{+0.070}_{-0.068}$ &$0.301^{+0.055}_{-0.059}$  &$0.043^{+0.049}_{-0.041}$  &$0.247^{+0.051}_{-0.051}$\\
$Age$			& Age (Gyr)			   &$7.2^{+4.0}_{-4.0}$			    &$8.1^{+2.7}_{-2.5}$		&$5.3^{+2.3}_{-1.9}$		&$3.9^{+0.8}_{-0.8}$\\
$A_V$			& $V$-band extinction (mag)    &$0.228^{+0.035}_{-0.064}$  &$0.119^{+0.021}_{-0.037}$  &$0.074^{+0.049}_{-0.045}$	&$0.180^{+0.049}_{-0.053}$\\
$d$				& Distance (pc)	       &$201.8^{+0. 6}_{-0.6}$	&$445.0^{+3.2}_{-3.2}$		&$525.4^{+2.8}_{-2.7}$		&$750.0^{+12.0}_{-12.0}$\\
Planetary Parameters:	& 	&	\\
$R_{\rm p}$			& Radius ($R_{\rm J}$)	&$1.008^{+0.014}_{-0.013}$		&$1.205^{+0.017}_{-0.016}$		&$1.324^{+0.023}_{-0.022}$			&$1.611^{+0.024}_{-0.024}$\\
$M_{\rm p}$			& Mass ($M_{\rm J}$)	&$0.277^{+0.017}_{-0.016}$		&$0.307^{+0.021}_{-0.020}$		&$0.596^{+0.073}_{-0.072}$	&$0.554^{+0.092}_{-0.091}$\\
$a$	& Semi-major axis (AU)    &$0.04599^{+0.00063}_{-0.00058}$  &$0.05042^{+0.00068}_{-0.00065}$    &$0.04628^{+0.00074}_{-0.00072}$    &$0.04042^{+0.00057}_{-0.00055}$\\
$T_{\rm eq}$		& Equilibrium temperature (K)	&$981.2^{+7.7}_{-8.1}$		&$1168.2^{+6.9}_{-7.0}$		&$1367.0^{+11.0}_{-11.0}$	&$1818.0^{+13.0}_{-13.0}$\\
$K$				& RV semi-amplitude ($\mathrm{m~s^{-1}}$)		&$40.9^{+2.2}_{-2.2}$		&$39.7^{+2.4}_{-2.5}$		&$77.6^{+9.1}_{-9.2}$	&$68.0^{+11.0}_{-11.0}$\\
$\rho_{\rm p}$		& Density (cgs)					&$0.335^{+0.019}_{-0.019}$	&$0.218^{+0.014}_{-0.014}$	&$0.318^{+0.040}_{-0.039}$	&$0.164^{+0.027}_{-0.027}$\\
${\rm log\ }g_{\rm p}$	& Surface gravity			&$2.830^{+0.023}_{-0.024}$	&$2.720^{+0.026}_{-0.028}$	&$2.926^{+0.050}_{-0.056}$	&$2.724^{+0.065}_{-0.077}$\\
$\Theta$		& Safronov Number	&$0.0313^{+0.0018}_{-0.0017}$	&$0.0267^{+0.0017}_{-0.0017}$	&$0.0405^{+0.0048}_{-0.0048}$			&$0.0214^{+0.0035}_{-0.0035}$\\
$<F>$			& Incident Flux ($\rm 10^9\ erg\ s^{-1}\ cm^{-2}$)	&$0.210^{+0.007}_{-0.007}$	&$0.422^{+0.010}_{-0.010}$	&$0.794^{+0.025}_{-0.024}$	& $2.478^{+0.071}_{-0.069}$\\
\hline
\end{tabular}}
\end{table*}

\begin{table}
\renewcommand\arraystretch{1.5}
\centering
\caption{Parameter estimation for spectral retrievals.}
\label{tab: retrieval params}
\scalebox{1.0}{
\begin{tabular}{cccc}
\hline\hline
Parameter & Prior & \multicolumn{2}{c}{Posterior}\\
 & & HAT-P-19b & HAT-P-65b\\
\hline
$T_\mathrm{iso}$ (K) & $\mathcal{U}(500,2500)$ & $681^{+118}_{-115}$ & $1412^{+140}_{-43}$\\
$R_\mathrm{10mbar}$ ($R_\mathrm{J}$) & $\mathcal{U}(0.5,2)$ & $1.032^{+0.005}_{-0.008}$ & $1.634^{+0.021}_{-0.024}$\\
$\log P_\mathrm{cloud}$ (bar) & $\mathcal{U}(-6,2)$ & $-2.3^{+2.8}_{-2.5}$ & $-1.4^{+2.2}_{-2.8}$\\
$\log A_\mathrm{scatt}$ & $\mathcal{U}(0,4)$ & $1.7^{+1.4}_{-1.2}$ & $2.1^{+1.3}_{-1.3}$\\
$\mathrm{C/O}$ & $\mathcal{U}(0.1,1.6)$ & $0.76^{+0.55}_{-0.45}$ & $0.94^{+0.38}_{-0.46}$\\
$\log Z/Z_\odot$ & $\mathcal{U}(-2,3)$ & $1.8^{+0.8}_{-1.1}$ & $0.3^{+1.4}_{-1.4}$\\
$\phi$ & $\mathcal{U}(0,1)$ & $0.38^{+0.37}_{-0.25}$ & $0.40^{+0.27}_{-0.24}$\\
Offset (ppm) & $\mathcal{U}(-5000,5000)$ & $1299^{+115}_{-118}$ & $-54^{+159}_{-164}$\\
\hline
\end{tabular}
}
\end{table}

\begin{figure*}
\centering
\includegraphics[width=\textwidth]{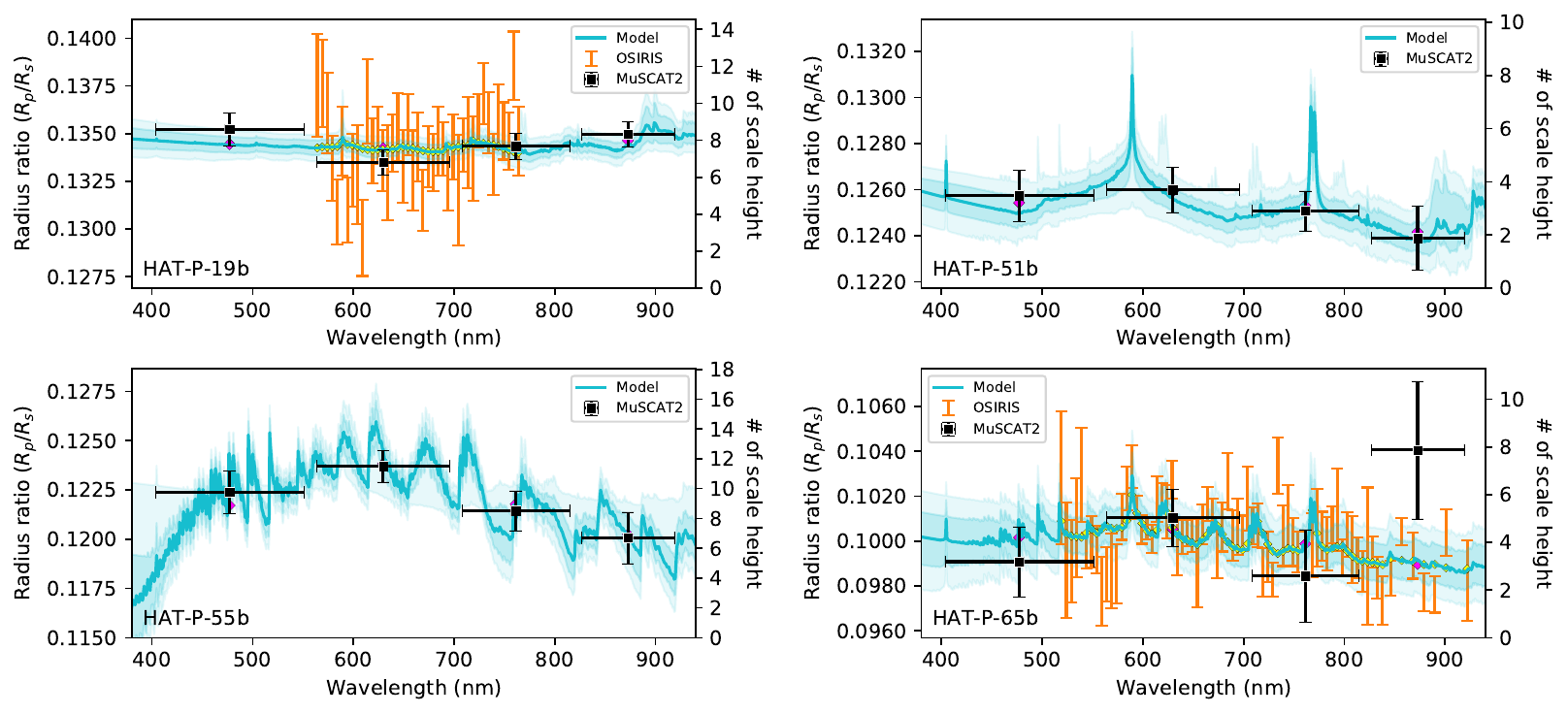}
\caption{Broadband transmission spectra of the four hot Jupiters (HAT-P-19b, HAT-P-51b, HAT-P-55b, and HAT-P-65b) along with the retrieved atmospheric models. For HAT-P-19b and HAT-P-65b, the retrievals were performed on the MuSCAT2 and OSIRIS combined dataset, with the assumption of patchy clouds and equilibrium chemistry. The OSIRIS data for HAT-P-19b and HAT-P-65b were obtained from \citet{2015AA...580A..60M} and \citet{2021ApJ...913L..16C}, respectively. For HAT-P-51b and HAT-P-55b, due to limited data points, the retrievals adopted a simplified assumption of cloud-free solar atmosphere.}
\label{fig: atmosphere}
\end{figure*}

\section{Variation of transit depth with wavelength}
\label{sec:transmission spectra}
Based on the MuSCAT2 data, we measured a difference between the maximum and minimum transit depths of $463\pm293$ ppm, $525\pm424$ ppm, $887\pm370$ ppm, and $1138\pm758$ ppm for HAT-P-19b, HAT-P-51b, HAT-P-55b, and HAT-P-65b, respectively, corresponding to $1.8\pm1.1$, $1.8\pm1.5$, $4.8\pm2.0$, and $5.4\pm3.6$ times the transit depth variation caused by one atmospheric scale height. The atmospheric scale height, $H=k_\mathrm{B}T_\mathrm{eq}/(\mu g_\mathrm{p})$, is estimated to be 0.0072, 0.0093, 0.0061, and 0.0107 $R_\mathrm{p}$, where $k_\mathrm{B}$ is the Boltzmann constant, $T_\mathrm{eq}$ is the planetary equilibrium temperature, $g_\mathrm{p}$ is the planetary surface gravity, and $\mu=2.3$~g\,mol$^{-1}$ is the mean molecular weight. 

The transit depths measured by MuSCAT2 in the $g$, $r$, $i$, $z_s$ bands sample a broadband transmission spectrum for each planet. In particular, MuSCAT2's ability to perform simultaneous multicolour photometry eliminates the impact of stellar rotational modulation, allowing us to take a first look at the planetary atmospheres. The optical transmission spectrum is sensitive to both optical absorbers (such as alkali metals and metal oxides) and particle sizes of scattering sources. However, the broad band averages the expected spectral features resulting from the planetary atmosphere, making it difficult to unambiguously distinguish the opacity sources of origin. Instead of using the broadband transmission spectra to infer atmospheric properties, we tried to answer which targets have a higher priority for follow-up transmission spectroscopy.

Assuming that the variation in transit depth is potentially caused by the planetary atmosphere, we performed a simplified Bayesian spectral retrieval analysis on the MuSCAT2 broadband transmission spectra. Two model hypotheses were considered, including a flat model and a planetary atmosphere model. The flat model has a constant planetary radius as the only free parameter. The planetary atmosphere model assumes a clear atmosphere of solar composition ($\mathrm{C/O}=0.55$, $\log Z/Z_\odot = 0$) in chemical equilibrium, consisting of two free parameters, the planetary radius at 10~mbar ($R_\mathrm{10mbar}$) and the isothermal temperature ($T_\mathrm{iso}$). We used petitRADTRANS \citep{2019A&A...627A..67M} to create the planetary atmosphere model, and PyMultiNest \citep{2014A&A...564A.125B} to implement the multimodal nested sampling \citep{2008MNRAS.384..449F,2009MNRAS.398.1601F} for parameter estimation. 

Figure \ref{fig: atmosphere} presents the MuSCAT2 broadband transmission spectrum along with the retrieved atmosphere models. Compared to the flat model, the atmosphere model resulted in decreasing reduced chi-square values for HAT-P-19b (from 1.18 to 1.08), HAT-P-51b (from 0.58 to 0.10), and HAT-P-55b (from 2.25 to 0.24), but an increasing value for HAT-P-65b (from 1.08 to 1.70), indicating that the atmosphere model provides a better fit than the flat model for the first three planets. We also calculated the log-evidence to compare these two models, and obtained $\Delta\ln\mathcal{Z}(=\ln\mathcal{Z}_\mathrm{atmos}-\ln\mathcal{Z}_\mathrm{flat})$ values of $-1.2\pm0.1$, $-0.2\pm0.1$, $2.5\pm0.1$, and $-0.5\pm0.1$ for HAT-P-19b, HAT-P-51b, HAT-P-55b, and HAT-P-65b, respectively. Therefore, HAT-P-55b is the only planet with moderate evidence in the Bayesian framework that a planetary atmosphere is required to explain the data, making it a priority target for future follow-up spectrophotometric observations. 

Of the four planets, two have been observed for optical transmission spectra prior to our MuSCAT2 observations. \citet{2015AA...580A..60M} obtained a flat featureless spectrum for HAT-P-19b using the R2500R grism of GTC's OSIRIS spectrograph, while \citet{2021ApJ...913L..16C} reported the detection of TiO and the possible detection of Na and VO in the atmosphere of HAT-P-65b using the R1000R grism of GTC OSIRIS. Therefore, we also performed retrievals on the combined MuSCAT2 and GTC dataset for HAT-P-19b and HAT-P-65b. In this case, we adopted a more complicated planetary atmosphere model because more data points were available. The model assumes an isothermal atmosphere at a temperature of $T_\mathrm{iso}$ in chemical equilibrium controlled by the metallicity $Z$ and the C/O ratio with a clear and a cloudy sector. The cloudy sector has a cloud fraction of $\phi$, a cloud top at pressure $P_\mathrm{cloud}$, and a scattering amplitude $A_\mathrm{scatt}$ times that of H$_2$ Rayleigh scattering. To account for the offsets introduced by different orbital parameters in deriving the transit depth and different instrumental systematics, the GTC OSIRIS spectra were allowed to have a free offset in the retrieval. 

Table \ref{tab: retrieval params} presents the retrieved parameters based on the combined MuSCAT2 and GTC dataset for HAT-P-19b and HAT-P-65b. For HAT-P-19b, the atmospheric metallicity tends to be super solar. For HAT-P-65b, the retrieved parameters agree well with those of \citet{2021ApJ...913L..16C}. Unfortunately, due to the lack of infrared wavelengths that cover sufficient molecular spectral features to characterise atmospheric chemistry and cloud altitude, it is difficult to constrain the parameters other than temperature, reference radius, and instrumental offset. Future transmission spectroscopy conducted with the James Webb Space Telescope, together with the current optical transmission spectra, should be able to place more meaningful constraints on the atmospheric metallicity, cloud properties, and relative elemental ratios, paving the way for tracing planetary formation and migration histories \citep[e.g.,][]{2011ApJ...743L..16O,2014ApJ...794L..12M,2016ApJ...832...41M,2021ApJ...914...12L,2023ApJ...946...18O}.

\section{Conclusions}
\label{sec:conclusions}
We performed simultaneous multicolour photometric observations of the transiting exoplanet systems HAT-P-19, HAT-P-51, HAT-P-55, and HAT-P-65 with the MuSCAT2 camera on the 1.52 m TCS telescope. We observed 12 transits for the four planets and obtained a total of 43 MuSCAT2 transit light curves. The transit parameters were revised based on the MuSCAT2 multicolour transit light curves. We also collected light curves for 56 transits from the TESS photometry, and combined the TESS timings with MuSCAT2 and literature timings to improve the orbital period and ephemeris estimates. We then consistently refined the physical parameters of these planetary systems by performing EXOFASTv2 global fits to the MuSCAT2 transit data, archival RV data, Gaia DR3 parallax, isochrones, and broadband spectral energy distributions. Finally, we investigated the potential for atmospheric characterisation using the MuSCAT2 multicolour transit depths for these four hot Jupiters. Our conclusions can be summarised as follows:
\begin{itemize}
    \item[-] We have improved the transit parameter estimates for HAT-P-19b, HAT-P-51b, and HAT-P-55b, with smaller uncertainties than previous studies. The MuSCAT2 uncertainties for HAT-P-65b are slightly larger than those derived from the very precise GTC observations.
    \item[-] We have consistently refined the physical parameters for all four planetary systems based on the improved transit parameters, which were derived from MuSCAT2 and GTC for HAT-P-65b, but only from MuSCAT2 for the other three. All the stellar and planetary radii are more tightly constrained than in previous studies, with typical relative errors of less than 2\%. 
    \item[-] We have improved the orbital period and ephemeris estimates for all four planetary systems. All of them are consistent with linear ephemeris. No significant transit timing variations or evidence of orbital decay were found. Based on our results, the typical uncertainties of the predicted mid-transit time by mid-2035 would be 33, 84, 77, and 104 seconds for HAT-P-19b, HAT-P-51b, HAT-P-55b, and HAT-P-65b, respectively, which are reasonably precise even in the ARIEL era.
    \item[-] We have found that planetary atmosphere models can improve the fit to the MuSCAT2 broadband transmission spectra of HAT-P-19b, HAT-P-51b, and HAT-P-55b compared to a flat line based on $\chi^2$ statistics. However, in terms of Bayesian model statistical significance, only HAT-P-55b shows (moderate) evidence of the presence of a planetary atmosphere. This makes HAT-P-55b a priority target for future transmission spectroscopy.
\end{itemize}

\section*{Acknowledgements}

    G.C. acknowledges the support by the B-type Strategic Priority Program of the Chinese Academy of Sciences (grant No. XDB41000000), the National Natural Science Foundation of China (NSFC; grant Nos. 42075122, 12122308), Youth Innovation Promotion Association CAS (2021315), and the Minor Planet Foundation of the Purple Mountain Observatory. 
    Y.M. acknowledges the support by NSFC (grant No. 12033010). 
    M.T. is supported by JSPS KAKENHI grant No.18H05442. 
    E.E-B. acknowledges financial support from the European Union and the State Agency of Investigation of the Spanish Ministry of Science and Innovation (MICINN) under the grant PRE2020-093107 of the Pre-Doc Program for the Training of Doctors (FPI-SO) through FSE funds. 
    R.L. acknowledges funding from University of La Laguna through the Margarita Salas Fellowship from the Spanish Ministry of Universities ref. UNI/551/2021-May 26, and under the EU Next Generation funds. 
    M.S. acknowledges the support of the Italian National Institute of Astrophysics (INAF) through the project 'The HOT-ATMOS Project: characterizing the atmospheres of hot giant planets as a key to understand the exoplanet diversity' (1.05.01.85.04). 
    This work is partly supported by JSPS KAKENHI Grant Numbers JP18H05439, JP21K13955, JP21K20376, and JST CREST Grant Number JPMJCR1761. 
    This article is based on observations made with the MuSCAT2 instrument, developed by ABC, at Telescopio Carlos S\'{a}nchez operated on the island of Tenerife by the IAC in the Spanish Observatorio del Teide. 
    This paper includes data collected with the TESS mission, obtained from the MAST data archive at the Space Telescope Science Institute (STScI). Funding for the TESS mission is provided by the NASA Explorer Program. STScI is operated by the Association of Universities for Research in Astronomy, Inc., under NASA contract NAS 5–26555. 
    This work has made use of data from the European Space Agency (ESA) mission {\it Gaia} (\url{https://www.cosmos.esa.int/gaia}), processed by the {\it Gaia} Data Processing and Analysis Consortium (DPAC, \url{https://www.cosmos.esa.int/web/gaia/dpac/consortium}). Funding for the DPAC has been provided by national institutions, in particular, the institutions participating in the {\it Gaia} Multilateral Agreement. 
    This work has made use of Matplotlib \citep{2007CSE.....9...90H}, the VizieR catalogue access tool, CDS, Strasbourg, France \citep{2000A&AS..143...23O}, and TEPCat \citep{2011MNRAS.417.2166S}.

\section*{Data Availability}
 
The data underlying this article will be shared at reasonable request to the corresponding author. The reduced light curves presented in this work will be made available at the CDS (\url{http://cdsarc.u-strasbg.fr/}).



\bibliographystyle{mnras}
\bibliography{ref} 

\begin{thebibliography}{}
\makeatletter
\relax
\def\mn@urlcharsother{\let\do\@makeother \do\$\do\&\do\#\do\^\do\_\do\%\do\~}
\def\mn@doi{\begingroup\mn@urlcharsother \@ifnextchar [ {\mn@doi@}
  {\mn@doi@[]}}
\def\mn@doi@[#1]#2{\def\@tempa{#1}\ifx\@tempa\@empty \href
  {http://dx.doi.org/#2} {doi:#2}\else \href {http://dx.doi.org/#2} {#1}\fi
  \endgroup}
\def\mn@eprint#1#2{\mn@eprint@#1:#2::\@nil}
\def\mn@eprint@arXiv#1{\href {http://arxiv.org/abs/#1} {{\tt arXiv:#1}}}
\def\mn@eprint@dblp#1{\href {http://dblp.uni-trier.de/rec/bibtex/#1.xml}
  {dblp:#1}}
\def\mn@eprint@#1:#2:#3:#4\@nil{\def\@tempa {#1}\def\@tempb {#2}\def\@tempc
  {#3}\ifx \@tempc \@empty \let \@tempc \@tempb \let \@tempb \@tempa \fi \ifx
  \@tempb \@empty \def\@tempb {arXiv}\fi \@ifundefined
  {mn@eprint@\@tempb}{\@tempb:\@tempc}{\expandafter \expandafter \csname
  mn@eprint@\@tempb\endcsname \expandafter{\@tempc}}}

\bibitem[\protect\citeauthoryear{{Ba{\c{s}}t{\"u}rk}, {Yal{\c{c}}{\i}nkaya},
  {Esmer}, {Tanr{\i}verdi}, {Mancini}, {Daylan}, {Southworth}  \&
  {Keten}}{{Ba{\c{s}}t{\"u}rk} et~al.}{2020}]{2020MNRAS.496.4174B}
{Ba{\c{s}}t{\"u}rk} {\"O}.,  {Yal{\c{c}}{\i}nkaya} S.,  {Esmer} E.~M.,
  {Tanr{\i}verdi} T.,  {Mancini} L.,  {Daylan} T.,  {Southworth} J.,   {Keten}
  B.,  2020, \mn@doi [\mnras] {10.1093/mnras/staa1758}, \href
  {https://ui.adsabs.harvard.edu/abs/2020MNRAS.496.4174B} {496, 4174}

\bibitem[\protect\citeauthoryear{{Buchner} et~al.,}{{Buchner}
  et~al.}{2014}]{2014A&A...564A.125B}
{Buchner} J.,  et~al., 2014, \mn@doi [\aap] {10.1051/0004-6361/201322971},
  \href {https://ui.adsabs.harvard.edu/abs/2014A&A...564A.125B} {564, A125}

\bibitem[\protect\citeauthoryear{{Charbonneau}, {Brown}, {Noyes}  \&
  {Gilliland}}{{Charbonneau} et~al.}{2002}]{2002ApJ...568..377C}
{Charbonneau} D.,  {Brown} T.~M.,  {Noyes} R.~W.,   {Gilliland} R.~L.,  2002,
  \mn@doi [\apj] {10.1086/338770}, \href
  {https://ui.adsabs.harvard.edu/abs/2002ApJ...568..377C} {568, 377}

\bibitem[\protect\citeauthoryear{{Chen} et~al.,}{{Chen}
  et~al.}{2014}]{2014A&A...563A..40C}
{Chen} G.,  et~al., 2014, \mn@doi [\aap] {10.1051/0004-6361/201322740}, \href
  {https://ui.adsabs.harvard.edu/abs/2014A&A...563A..40C} {563, A40}

\bibitem[\protect\citeauthoryear{{Chen} et~al.,}{{Chen}
  et~al.}{2021a}]{2021MNRAS.500.5420C}
{Chen} G.,  et~al., 2021a, \mn@doi [\mnras] {10.1093/mnras/staa3555}, \href
  {https://ui.adsabs.harvard.edu/abs/2021MNRAS.500.5420C} {500, 5420}

\bibitem[\protect\citeauthoryear{{Chen}, {Pall{\'e}}, {Parviainen}, {Murgas}
  \& {Yan}}{{Chen} et~al.}{2021b}]{2021ApJ...913L..16C}
{Chen} G.,  {Pall{\'e}} E.,  {Parviainen} H.,  {Murgas} F.,   {Yan} F.,  2021b,
  \mn@doi [\apjl] {10.3847/2041-8213/abfbe1}, \href
  {https://ui.adsabs.harvard.edu/abs/2021ApJ...913L..16C} {913, L16}

\bibitem[\protect\citeauthoryear{{Cutri} \& {et al.}}{{Cutri} \& {et
  al.}}{2014}]{2014yCat.2328....0C}
{Cutri} R.~M.,  {et al.} 2014, VizieR Online Data Catalog, \href
  {https://ui.adsabs.harvard.edu/abs/2014yCat.2328....0C} {p. II/328}

\bibitem[\protect\citeauthoryear{{Dotter}}{{Dotter}}{2016}]{2016ApJS..222....8D}
{Dotter} A.,  2016, \mn@doi [\apjs] {10.3847/0067-0049/222/1/8}, \href
  {https://ui.adsabs.harvard.edu/abs/2016ApJS..222....8D} {222, 8}

\bibitem[\protect\citeauthoryear{{Droege}, {Richmond}, {Sallman}  \&
  {Creager}}{{Droege} et~al.}{2006}]{2006PASP..118.1666D}
{Droege} T.~F.,  {Richmond} M.~W.,  {Sallman} M.~P.,   {Creager} R.~P.,  2006,
  \mn@doi [\pasp] {10.1086/510197}, \href
  {https://ui.adsabs.harvard.edu/abs/2006PASP..118.1666D} {118, 1666}

\bibitem[\protect\citeauthoryear{{Eastman}, {Siverd}  \& {Gaudi}}{{Eastman}
  et~al.}{2010}]{2010PASP..122..935E}
{Eastman} J.,  {Siverd} R.,   {Gaudi} B.~S.,  2010, \mn@doi [\pasp]
  {10.1086/655938}, \href
  {https://ui.adsabs.harvard.edu/abs/2010PASP..122..935E} {122, 935}

\bibitem[\protect\citeauthoryear{{Eastman} et~al.,}{{Eastman}
  et~al.}{2019}]{2019arXiv190709480E}
{Eastman} J.~D.,  et~al., 2019, arXiv e-prints, \href
  {https://ui.adsabs.harvard.edu/abs/2019arXiv190709480E} {p. arXiv:1907.09480}

\bibitem[\protect\citeauthoryear{{Edwards} et~al.,}{{Edwards}
  et~al.}{2021}]{2021MNRAS.504.5671E}
{Edwards} B.,  et~al., 2021, \mn@doi [\mnras] {10.1093/mnras/staa1245}, \href
  {https://ui.adsabs.harvard.edu/abs/2021MNRAS.504.5671E} {504, 5671}

\bibitem[\protect\citeauthoryear{{Espinoza} \& {Jord{\'a}n}}{{Espinoza} \&
  {Jord{\'a}n}}{2015}]{2015MNRAS.450.1879E}
{Espinoza} N.,  {Jord{\'a}n} A.,  2015, \mn@doi [\mnras]
  {10.1093/mnras/stv744}, \href
  {https://ui.adsabs.harvard.edu/abs/2015MNRAS.450.1879E} {450, 1879}

\bibitem[\protect\citeauthoryear{{Feroz} \& {Hobson}}{{Feroz} \&
  {Hobson}}{2008}]{2008MNRAS.384..449F}
{Feroz} F.,  {Hobson} M.~P.,  2008, \mn@doi [\mnras]
  {10.1111/j.1365-2966.2007.12353.x}, \href
  {https://ui.adsabs.harvard.edu/abs/2008MNRAS.384..449F} {384, 449}

\bibitem[\protect\citeauthoryear{{Feroz}, {Hobson}  \& {Bridges}}{{Feroz}
  et~al.}{2009}]{2009MNRAS.398.1601F}
{Feroz} F.,  {Hobson} M.~P.,   {Bridges} M.,  2009, \mn@doi [\mnras]
  {10.1111/j.1365-2966.2009.14548.x}, \href
  {https://ui.adsabs.harvard.edu/abs/2009MNRAS.398.1601F} {398, 1601}

\bibitem[\protect\citeauthoryear{{Foreman-Mackey} et~al.,}{{Foreman-Mackey}
  et~al.}{2013}]{2013ascl.soft03002F}
{Foreman-Mackey} D.,  et~al., 2013, {emcee: The MCMC Hammer}, Astrophysics
  Source Code Library, record ascl:1303.002 (\mn@eprint {ascl} {1303.002})

\bibitem[\protect\citeauthoryear{{Foreman-Mackey}, {Agol}, {Ambikasaran}  \&
  {Angus}}{{Foreman-Mackey} et~al.}{2017}]{2017AJ....154..220F}
{Foreman-Mackey} D.,  {Agol} E.,  {Ambikasaran} S.,   {Angus} R.,  2017,
  \mn@doi [\aj] {10.3847/1538-3881/aa9332}, \href
  {https://ui.adsabs.harvard.edu/abs/2017AJ....154..220F} {154, 220}

\bibitem[\protect\citeauthoryear{{Fortney}, {Marley}  \& {Barnes}}{{Fortney}
  et~al.}{2007}]{2007ApJ...659.1661F}
{Fortney} J.~J.,  {Marley} M.~S.,   {Barnes} J.~W.,  2007, \mn@doi [\apj]
  {10.1086/512120}, \href
  {https://ui.adsabs.harvard.edu/abs/2007ApJ...659.1661F} {659, 1661}

\bibitem[\protect\citeauthoryear{{Gaia Collaboration}}{{Gaia
  Collaboration}}{2022}]{2022yCat.1355....0G}
{Gaia Collaboration} 2022, VizieR Online Data Catalog, \href
  {https://ui.adsabs.harvard.edu/abs/2022yCat.1355....0G} {p. I/355}

\bibitem[\protect\citeauthoryear{{Gibson}, {Aigrain}, {Roberts}, {Evans},
  {Osborne}  \& {Pont}}{{Gibson} et~al.}{2012}]{2012MNRAS.419.2683G}
{Gibson} N.~P.,  {Aigrain} S.,  {Roberts} S.,  {Evans} T.~M.,  {Osborne} M.,
  {Pont} F.,  2012, \mn@doi [\mnras] {10.1111/j.1365-2966.2011.19915.x}, \href
  {https://ui.adsabs.harvard.edu/abs/2012MNRAS.419.2683G} {419, 2683}

\bibitem[\protect\citeauthoryear{{Greiner} et~al.,}{{Greiner}
  et~al.}{2008}]{2008PASP..120..405G}
{Greiner} J.,  et~al., 2008, \mn@doi [\pasp] {10.1086/587032}, \href
  {https://ui.adsabs.harvard.edu/abs/2008PASP..120..405G} {120, 405}

\bibitem[\protect\citeauthoryear{{Hagey}, {Edwards}  \& {Boley}}{{Hagey}
  et~al.}{2022}]{2022AJ....164..220H}
{Hagey} S.~R.,  {Edwards} B.,   {Boley} A.~C.,  2022, \mn@doi [\aj]
  {10.3847/1538-3881/ac959a}, \href
  {https://ui.adsabs.harvard.edu/abs/2022AJ....164..220H} {164, 220}

\bibitem[\protect\citeauthoryear{{Hartman} et~al.,}{{Hartman}
  et~al.}{2011}]{2011ApJ...726...52H}
{Hartman} J.~D.,  et~al., 2011, \mn@doi [\apj] {10.1088/0004-637X/726/1/52},
  \href {https://ui.adsabs.harvard.edu/abs/2011ApJ...726...52H} {726, 52}

\bibitem[\protect\citeauthoryear{{Hartman} et~al.,}{{Hartman}
  et~al.}{2015}]{2015AJ....150..168H}
{Hartman} J.~D.,  et~al., 2015, \mn@doi [\aj] {10.1088/0004-6256/150/6/168},
  \href {https://ui.adsabs.harvard.edu/abs/2015AJ....150..168H} {150, 168}

\bibitem[\protect\citeauthoryear{{Hartman} et~al.,}{{Hartman}
  et~al.}{2016}]{2016AJ....152..182H}
{Hartman} J.~D.,  et~al., 2016, \mn@doi [\aj] {10.3847/0004-6256/152/6/182},
  \href {https://ui.adsabs.harvard.edu/abs/2016AJ....152..182H} {152, 182}

\bibitem[\protect\citeauthoryear{{Henden}, {Templeton}, {Terrell}, {Smith},
  {Levine}  \& {Welch}}{{Henden} et~al.}{2016}]{2016yCat.2336....0H}
{Henden} A.~A.,  {Templeton} M.,  {Terrell} D.,  {Smith} T.~C.,  {Levine} S.,
  {Welch} D.,  2016, VizieR Online Data Catalog, \href
  {https://ui.adsabs.harvard.edu/abs/2016yCat.2336....0H} {p. II/336}

\bibitem[\protect\citeauthoryear{{Hunter}}{{Hunter}}{2007}]{2007CSE.....9...90H}
{Hunter} J.~D.,  2007, \mn@doi [Computing in Science and Engineering]
  {10.1109/MCSE.2007.55}, \href
  {https://ui.adsabs.harvard.edu/abs/2007CSE.....9...90H} {9, 90}

\bibitem[\protect\citeauthoryear{{Ivshina} \& {Winn}}{{Ivshina} \&
  {Winn}}{2022}]{2022ApJS..259...62I}
{Ivshina} E.~S.,  {Winn} J.~N.,  2022, \mn@doi [\apjs]
  {10.3847/1538-4365/ac545b}, \href
  {https://ui.adsabs.harvard.edu/abs/2022ApJS..259...62I} {259, 62}

\bibitem[\protect\citeauthoryear{{Jenkins} et~al.,}{{Jenkins}
  et~al.}{2016}]{2016SPIE.9913E..3EJ}
{Jenkins} J.~M.,  et~al., 2016, in {Chiozzi} G.,  {Guzman} J.~C.,  eds,
  Society of Photo-Optical Instrumentation Engineers (SPIE) Conference Series
  Vol. 9913, Software and Cyberinfrastructure for Astronomy IV. p. 99133E,
  \mn@doi{10.1117/12.2233418}

\bibitem[\protect\citeauthoryear{{Juncher} et~al.,}{{Juncher}
  et~al.}{2015}]{2015PASP..127..851J}
{Juncher} D.,  et~al., 2015, \mn@doi [\pasp] {10.1086/682725}, \href
  {https://ui.adsabs.harvard.edu/abs/2015PASP..127..851J} {127, 851}

\bibitem[\protect\citeauthoryear{{Kipping}}{{Kipping}}{2010}]{2010MNRAS.408.1758K}
{Kipping} D.~M.,  2010, \mn@doi [\mnras] {10.1111/j.1365-2966.2010.17242.x},
  \href {https://ui.adsabs.harvard.edu/abs/2010MNRAS.408.1758K} {408, 1758}

\bibitem[\protect\citeauthoryear{{Kokori} et~al.,}{{Kokori}
  et~al.}{2022}]{2022ApJS..258...40K}
{Kokori} A.,  et~al., 2022, \mn@doi [\apjs] {10.3847/1538-4365/ac3a10}, \href
  {https://ui.adsabs.harvard.edu/abs/2022ApJS..258...40K} {258, 40}

\bibitem[\protect\citeauthoryear{{Kreidberg}}{{Kreidberg}}{2015}]{2015ascl.soft10002K}
{Kreidberg} L.,  2015, {batman: BAsic Transit Model cAlculatioN in Python},
  Astrophysics Source Code Library, record ascl:1510.002 (\mn@eprint {ascl}
  {1510.002})

\bibitem[\protect\citeauthoryear{{Lightkurve Collaboration}
  et~al.,}{{Lightkurve Collaboration} et~al.}{2018}]{2018ascl.soft12013L}
{Lightkurve Collaboration} et~al., 2018, {Lightkurve: Kepler and TESS time
  series analysis in Python} (\mn@eprint {ascl} {1812.013})

\bibitem[\protect\citeauthoryear{{Lothringer}, {Rustamkulov}, {Sing}, {Gibson},
  {Wilson}  \& {Schlaufman}}{{Lothringer} et~al.}{2021}]{2021ApJ...914...12L}
{Lothringer} J.~D.,  {Rustamkulov} Z.,  {Sing} D.~K.,  {Gibson} N.~P.,
  {Wilson} J.,   {Schlaufman} K.~C.,  2021, \mn@doi [\apj]
  {10.3847/1538-4357/abf8a9}, \href
  {https://ui.adsabs.harvard.edu/abs/2021ApJ...914...12L} {914, 12}

\bibitem[\protect\citeauthoryear{{Luque} et~al.,}{{Luque}
  et~al.}{2020}]{2020A&A...642A..50L}
{Luque} R.,  et~al., 2020, \mn@doi [\aap] {10.1051/0004-6361/202038703}, \href
  {https://ui.adsabs.harvard.edu/abs/2020A&A...642A..50L} {642, A50}

\bibitem[\protect\citeauthoryear{{Maciejewski}, {Stangret}, {Ohlert},
  {Basaran}, {Maciejczak}, {Puciata-Mroczynska}  \& {Boulanger}}{{Maciejewski}
  et~al.}{2018}]{2018IBVS.6243....1M}
{Maciejewski} G.,  {Stangret} M.,  {Ohlert} J.,  {Basaran} C.~S.,  {Maciejczak}
  J.,  {Puciata-Mroczynska} M.,   {Boulanger} E.,  2018, \mn@doi [Information
  Bulletin on Variable Stars] {10.22444/IBVS.6243}, \href
  {https://ui.adsabs.harvard.edu/abs/2018IBVS.6243....1M} {6243, 1}

\bibitem[\protect\citeauthoryear{{Madhusudhan}, {Amin}  \&
  {Kennedy}}{{Madhusudhan} et~al.}{2014}]{2014ApJ...794L..12M}
{Madhusudhan} N.,  {Amin} M.~A.,   {Kennedy} G.~M.,  2014, \mn@doi [\apjl]
  {10.1088/2041-8205/794/1/L12}, \href
  {https://ui.adsabs.harvard.edu/abs/2014ApJ...794L..12M} {794, L12}

\bibitem[\protect\citeauthoryear{{Madhusudhan}, {Ag{\'u}ndez}, {Moses}  \&
  {Hu}}{{Madhusudhan} et~al.}{2016}]{2016SSRv..205..285M}
{Madhusudhan} N.,  {Ag{\'u}ndez} M.,  {Moses} J.~I.,   {Hu} Y.,  2016, \mn@doi
  [\ssr] {10.1007/s11214-016-0254-3}, \href
  {https://ui.adsabs.harvard.edu/abs/2016SSRv..205..285M} {205, 285}

\bibitem[\protect\citeauthoryear{{Mallonn} et~al.,}{{Mallonn}
  et~al.}{2015}]{2015AA...580A..60M}
{Mallonn} M.,  et~al., 2015, \mn@doi [\aap] {10.1051/0004-6361/201423778},
  \href {https://ui.adsabs.harvard.edu/abs/2015A&A...580A..60M} {580, A60}

\bibitem[\protect\citeauthoryear{{Mancini} et~al.,}{{Mancini}
  et~al.}{2014}]{2014MNRAS.443.2391M}
{Mancini} L.,  et~al., 2014, \mn@doi [\mnras] {10.1093/mnras/stu1286}, \href
  {https://ui.adsabs.harvard.edu/abs/2014MNRAS.443.2391M} {443, 2391}

\bibitem[\protect\citeauthoryear{{Mandel} \& {Agol}}{{Mandel} \&
  {Agol}}{2002}]{2002ApJ...580L.171M}
{Mandel} K.,  {Agol} E.,  2002, \mn@doi [\apjl] {10.1086/345520}, \href
  {https://ui.adsabs.harvard.edu/abs/2002ApJ...580L.171M} {580, L171}

\bibitem[\protect\citeauthoryear{{Mayor} \& {Queloz}}{{Mayor} \&
  {Queloz}}{1995}]{1995Natur.378..355M}
{Mayor} M.,  {Queloz} D.,  1995, \mn@doi [\nat] {10.1038/378355a0}, \href
  {https://ui.adsabs.harvard.edu/abs/1995Natur.378..355M} {378, 355}

\bibitem[\protect\citeauthoryear{{Medan}, {L{\'e}pine}  \& {Hartman}}{{Medan}
  et~al.}{2021}]{2021AJ....161..234M}
{Medan} I.,  {L{\'e}pine} S.,   {Hartman} Z.,  2021, \mn@doi [\aj]
  {10.3847/1538-3881/abe878}, \href
  {https://ui.adsabs.harvard.edu/abs/2021AJ....161..234M} {161, 234}

\bibitem[\protect\citeauthoryear{{Molli{\`e}re}, {Wardenier}, {van Boekel},
  {Henning}, {Molaverdikhani}  \& {Snellen}}{{Molli{\`e}re}
  et~al.}{2019}]{2019A&A...627A..67M}
{Molli{\`e}re} P.,  {Wardenier} J.~P.,  {van Boekel} R.,  {Henning} T.,
  {Molaverdikhani} K.,   {Snellen} I.~A.~G.,  2019, \mn@doi [\aap]
  {10.1051/0004-6361/201935470}, \href
  {https://ui.adsabs.harvard.edu/abs/2019A&A...627A..67M} {627, A67}

\bibitem[\protect\citeauthoryear{{Mordasini}, {van Boekel}, {Molli{\`e}re},
  {Henning}  \& {Benneke}}{{Mordasini} et~al.}{2016}]{2016ApJ...832...41M}
{Mordasini} C.,  {van Boekel} R.,  {Molli{\`e}re} P.,  {Henning} T.,
  {Benneke} B.,  2016, \mn@doi [\apj] {10.3847/0004-637X/832/1/41}, \href
  {https://ui.adsabs.harvard.edu/abs/2016ApJ...832...41M} {832, 41}

\bibitem[\protect\citeauthoryear{{Narita} et~al.,}{{Narita}
  et~al.}{2015}]{2015JATIS...1d5001N}
{Narita} N.,  et~al., 2015, \mn@doi [Journal of Astronomical Telescopes,
  Instruments, and Systems] {10.1117/1.JATIS.1.4.045001}, \href
  {https://ui.adsabs.harvard.edu/abs/2015JATIS...1d5001N} {1, 045001}

\bibitem[\protect\citeauthoryear{{Narita} et~al.,}{{Narita}
  et~al.}{2019}]{2019JATIS...5a5001N}
{Narita} N.,  et~al., 2019, \mn@doi [Journal of Astronomical Telescopes,
  Instruments, and Systems] {10.1117/1.JATIS.5.1.015001}, \href
  {https://ui.adsabs.harvard.edu/abs/2019JATIS...5a5001N} {5, 015001}

\bibitem[\protect\citeauthoryear{{Narita} et~al.,}{{Narita}
  et~al.}{2020}]{2020SPIE11447E..5KN}
{Narita} N.,  et~al., 2020, in Society of Photo-Optical Instrumentation
  Engineers (SPIE) Conference Series. p. 114475K, \mn@doi{10.1117/12.2559947}

\bibitem[\protect\citeauthoryear{{{\"O}berg}, {Murray-Clay}  \&
  {Bergin}}{{{\"O}berg} et~al.}{2011}]{2011ApJ...743L..16O}
{{\"O}berg} K.~I.,  {Murray-Clay} R.,   {Bergin} E.~A.,  2011, \mn@doi [\apjl]
  {10.1088/2041-8205/743/1/L16}, \href
  {https://ui.adsabs.harvard.edu/abs/2011ApJ...743L..16O} {743, L16}

\bibitem[\protect\citeauthoryear{{Ochsenbein}, {Bauer}  \&
  {Marcout}}{{Ochsenbein} et~al.}{2000}]{2000A&AS..143...23O}
{Ochsenbein} F.,  {Bauer} P.,   {Marcout} J.,  2000, \mn@doi [\aaps]
  {10.1051/aas:2000169}, \href
  {https://ui.adsabs.harvard.edu/abs/2000A&AS..143...23O} {143, 23}

\bibitem[\protect\citeauthoryear{{Ohno} \& {Fortney}}{{Ohno} \&
  {Fortney}}{2023}]{2023ApJ...946...18O}
{Ohno} K.,  {Fortney} J.~J.,  2023, \mn@doi [\apj] {10.3847/1538-4357/acafed},
  \href {https://ui.adsabs.harvard.edu/abs/2023ApJ...946...18O} {946, 18}

\bibitem[\protect\citeauthoryear{{Parviainen} et~al.,}{{Parviainen}
  et~al.}{2019}]{2019A&A...630A..89P}
{Parviainen} H.,  et~al., 2019, \mn@doi [\aap] {10.1051/0004-6361/201935709},
  \href {https://ui.adsabs.harvard.edu/abs/2019A&A...630A..89P} {630, A89}

\bibitem[\protect\citeauthoryear{{Ricker} et~al.,}{{Ricker}
  et~al.}{2015}]{2015JATIS...1a4003R}
{Ricker} G.~R.,  et~al., 2015, \mn@doi [Journal of Astronomical Telescopes,
  Instruments, and Systems] {10.1117/1.JATIS.1.1.014003}, \href
  {https://ui.adsabs.harvard.edu/abs/2015JATIS...1a4003R} {1, 014003}

\bibitem[\protect\citeauthoryear{{Seager} \& {Mall{\'e}n-Ornelas}}{{Seager} \&
  {Mall{\'e}n-Ornelas}}{2003}]{2003ApJ...585.1038S}
{Seager} S.,  {Mall{\'e}n-Ornelas} G.,  2003, \mn@doi [\apj] {10.1086/346105},
  \href {https://ui.adsabs.harvard.edu/abs/2003ApJ...585.1038S} {585, 1038}

\bibitem[\protect\citeauthoryear{{Seager} \& {Sasselov}}{{Seager} \&
  {Sasselov}}{2000}]{2000ApJ...537..916S}
{Seager} S.,  {Sasselov} D.~D.,  2000, \mn@doi [\apj] {10.1086/309088}, \href
  {https://ui.adsabs.harvard.edu/abs/2000ApJ...537..916S} {537, 916}

\bibitem[\protect\citeauthoryear{{Seeliger} et~al.,}{{Seeliger}
  et~al.}{2015}]{2015MNRAS.451.4060S}
{Seeliger} M.,  et~al., 2015, \mn@doi [\mnras] {10.1093/mnras/stv1187}, \href
  {https://ui.adsabs.harvard.edu/abs/2015MNRAS.451.4060S} {451, 4060}

\bibitem[\protect\citeauthoryear{{Southworth}}{{Southworth}}{2011}]{2011MNRAS.417.2166S}
{Southworth} J.,  2011, \mn@doi [\mnras] {10.1111/j.1365-2966.2011.19399.x},
  \href {https://ui.adsabs.harvard.edu/abs/2011MNRAS.417.2166S} {417, 2166}

\bibitem[\protect\citeauthoryear{{Yeh}, {Jiang}  \& {A-thano}}{{Yeh}
  et~al.}{2024}]{2024NewA..10602130Y}
{Yeh} L.-C.,  {Jiang} I.-G.,   {A-thano} N.,  2024, \mn@doi [\na]
  {10.1016/j.newast.2023.102130}, \href
  {https://ui.adsabs.harvard.edu/abs/2024NewA..10602130Y} {106, 102130}

\bibitem[\protect\citeauthoryear{{Zacharias}, {Finch}  \&
  {Frouard}}{{Zacharias} et~al.}{2017}]{2017yCat.1340....0Z}
{Zacharias} N.,  {Finch} C.,   {Frouard} J.,  2017, VizieR Online Data Catalog,
  \href {https://ui.adsabs.harvard.edu/abs/2017yCat.1340....0Z} {p. I/340}

\makeatother
\end{thebibliography}




\appendix
\label{sec:appendix}

\section{Additional tables and figures}

\begin{table*}
\renewcommand\arraystretch{1.5}
\centering
\caption{Adopted priors for limb-darkening coefficients.}
\label{tab: ldc priors}
\scalebox{0.9}{
\begin{tabular}{cccc}
\hline\hline
System	  & Band     & $u_1$   & $u_2$\\
\hline
HAT-P-19  &  $g$   & $\mathcal{N}(0.8106,0.0743)$ &  $\mathcal{N}(0.0221,0.0659)$\\
HAT-P-19  &  $r$   & $\mathcal{N}(0.5636,0.0560)$ &  $\mathcal{N}(0.1680,0.0410)$\\
HAT-P-19  &  $i$   & $\mathcal{N}(0.4446,0.0408)$ &  $\mathcal{N}(0.1993,0.0263)$\\
HAT-P-19  &  $z_s$ & $\mathcal{N}(0.3714,0.0342)$ &  $\mathcal{N}(0.2144,0.0202)$\\
\hline
HAT-P-51  &  $g$   & $\mathcal{N}(0.6814,0.0672)$ &  $\mathcal{N}(0.1280,0.0538)$\\
HAT-P-51  &  $r$   & $\mathcal{N}(0.4614,0.0533)$ &  $\mathcal{N}(0.2385,0.0346)$\\
HAT-P-51  &  $i$   & $\mathcal{N}(0.3665,0.0419)$ &  $\mathcal{N}(0.2506,0.0255)$\\
HAT-P-51  &  $z_s$ & $\mathcal{N}(0.3051,0.0351)$ &  $\mathcal{N}(0.2546,0.0192)$\\
\hline
HAT-P-55  &  $g$   & $\mathcal{N}(0.5792,0.0640)$ &  $\mathcal{N}(0.2048,0.0472)$\\
HAT-P-55  &  $r$   & $\mathcal{N}(0.3883,0.0479)$ &  $\mathcal{N}(0.2781,0.0277)$\\
HAT-P-55  &  $i$   & $\mathcal{N}(0.3104,0.0390)$ &  $\mathcal{N}(0.2745,0.0206)$\\
HAT-P-55  &  $z_s$ & $\mathcal{N}(0.2602,0.0345)$ &  $\mathcal{N}(0.2725,0.0171)$\\
\hline
HAT-P-65  &  $g$   & $\mathcal{N}(0.5898,0.0630)$ &  $\mathcal{N}(0.1966,0.0464)$\\
HAT-P-65  &  $r$   & $\mathcal{N}(0.3886,0.0496)$ &  $\mathcal{N}(0.2800,0.0292)$\\
HAT-P-65  &  $i$   & $\mathcal{N}(0.3064,0.0414)$ &  $\mathcal{N}(0.2807,0.0225)$\\
HAT-P-65  &  $z_s$ & $\mathcal{N}(0.2548,0.0358)$ &  $\mathcal{N}(0.2774,0.0182)$\\
\hline
\end{tabular}
}
\end{table*}

\begin{table*}
\renewcommand\arraystretch{1.5}
\centering
\caption{Broadband photometry and stellar parallax of HAT-P-19, HAT-P-51, HAT-P-55 and HAT-P-65.}
\label{tab: brightness of four hot Jupiters}
\scalebox{0.9}{
\begin{tabular}{ccccccccc}
\hline\hline
Passband	    & HAT-P-19 (mag)     & Ref.	 & HAT-P-51 (mag)     & Ref. &HAT-P-55 (mag)   & Ref.	& HAT-P-65 (mag)   & Ref.\\
\hline
Johnson $B$ 	& $13.834\pm0.051$   & 1	 & $14.261\pm0.067$	  & 1	 & $13.871\pm0.039$  & 1   & $13.818\pm0.021$   & 1\\
Johnson $V$ 	& $12.853\pm0.055$	 & 1     & $13.440\pm0.042$	  & 1	 & $13.207\pm0.039$  & 1  & $13.145\pm0.029$    & 1\\
Johnson $R$     & $11.990\pm0.100$   & 2     &  \text{-}  & \text{-}   &  \text{-}  & \text{-}  &  \text{-}  & \text{-}\\
Johnson $I$ 	&  \text{-}  & \text{-}  &  \text{-}  & \text{-}   &  \text{-}  & \text{-}   & $12.456\pm0.101$  & 6\\
SDSS $u^{'}$ 	& $15.589\pm0.005$	 & 3     &  \text{-}  & \text{-}   &  \text{-}  & \text{-}  &  \text{-}  & \text{-} \\
SDSS $g^{'}$ 	& $13.779\pm0.003$	 & 3     & $13.839\pm0.050$   & 1  & $13.501\pm0.039$ 	& 1 & $13.445\pm0.016$	& 3\\
SDSS $r^{'}$ 	& $12.659\pm0.002$	 & 3     & $13.194\pm0.032$   & 1  & $13.060\pm0.021$ 	& 1 & $12.948\pm0.033$	& 3\\
SDSS $i^{'}$ 	& $12.406\pm0.001$	 & 3     & $12.998\pm0.042$   & 1  & $12.880\pm0.041$ 	& 1 & $12.784\pm0.097$	& 3\\
SDSS $z^{'}$ 	& $12.623\pm0.004$	 & 3     & \text{-}   & \text{-}   &  \text{-}  & \text{-}   &  \text{-}  & \text{-}\\
2MASS $J$ 	    & $11.095\pm0.020$	 & 3    & $12.039\pm0.022$	& 5	& $12.020\pm0.022$	& 5	& $11.892\pm0.026$ & 5\\
2MASS $H$ 	    & $10.644\pm0.022$	 & 3    & $11.645\pm0.023$	& 5	& $11.714\pm0.026$	& 5	& $11.604\pm0.022$ & 5\\
2MASS $K_s$ 	& $10.546\pm0.019$	 & 3    & $11.614\pm0.020$	& 5	& $11.627\pm0.025$	& 5	& $11.528\pm0.025$ & 5\\
WISE1 	& $10.495\pm0.022$	& 3  &  \text{-}  & \text{-}   &  \text{-}  & \text{-} & $11.494\pm0.023$   & 5\\
WISE2 	& $10.557\pm0.020$	& 3  &  \text{-}  & \text{-}   &  \text{-}  & \text{-} & $11.532\pm0.021$   & 5\\
WISE3 	& $10.561\pm0.091$	& 3  &  \text{-}  & \text{-}   &  \text{-}  & \text{-} & $11.373\pm0.172$   & 5\\
Gaia    & $12.546\pm0.003$  & 4  & $13.271\pm0.003$   & 4  & $13.083\pm0.003$   & 4   & $12.981\pm0.003$   & 4\\
Gaia BP  & $13.059\pm0.003$  & 4  & $13.686\pm0.003$   & 4  & $13.414\pm0.003$   & 4   & $13.332\pm0.003$     & 4\\
Gaia RP  & $11.884\pm0.004$  & 4  & $12.698\pm0.004$   & 4  & $12.590\pm0.004$   & 4   & $12.468\pm0.004$   & 4\\
\hline
Parallax (mas) & $4.96\pm0.01$ & 4 & $2.25\pm0.02$ & 4 & $1.90\pm0.01$ & 4 & $1.32\pm0.03$ & 4\\
\hline
\end{tabular}
}
\begin{flushleft}
    \noindent{\footnotesize{References.(1) \citet{2016yCat.2336....0H}. (2) \citet{2017yCat.1340....0Z}. (3) \citet{2021AJ....161..234M}.(4) \citet{2022yCat.1355....0G}. (5) \citet{2014yCat.2328....0C}. (6) \citet{2006PASP..118.1666D}.}}
\end{flushleft}
\end{table*}


\begin{table}
\renewcommand\arraystretch{1.5}
\centering
\caption{Mid-transit times of the four hot Jupiters.}
\label{tab: mid-transit time}
\scalebox{0.8}{
\begin{tabular}{cccc}
\hline\hline
Planet     & Telescope       & $T_{\rm mid} - 2450000 [\rm BJD_{TDB}]$ & Ref.	\\
\hline
HAT-P-19b  &    TCS          & $8322.61426\pm0.00016$ & 1\\
HAT-P-19b  &    TCS          & $8334.64066\pm0.00010$ & 1\\
HAT-P-19b  &    TCS          & $8338.64963\pm0.00012$ & 1\\
HAT-P-19b  &    TESS         & $8767.58920\pm0.00089$ & 1\\
HAT-P-19b  &    TESS         & $8771.60119\pm0.00135$ & 1\\
HAT-P-19b  &    TESS         & $8775.60828\pm0.00126$ & 1\\
HAT-P-19b  &    TESS         & $8779.61650\pm0.00115$ & 1\\
HAT-P-19b  &    TESS         & $8783.62458\pm0.00097$ & 1\\
HAT-P-19b  &    TESS         & $8787.63334\pm0.00109$ & 1\\
HAT-P-19b  &    TESS         & $9853.96840\pm0.00153$ & 1\\
HAT-P-19b  &    TESS         & $9857.97795\pm0.00112$ & 1\\
HAT-P-19b  &    TESS         & $9861.98675\pm0.00077$ & 1\\
HAT-P-19b  &    TESS         & $9870.00479\pm0.00103$ & 1\\
HAT-P-19b  &    TESS         & $9874.01519\pm0.00082$ & 1\\
HAT-P-19b  &    TESS         & $9878.02177\pm0.00132$ & 1\\
HAT-P-19b  &    TESS         & $9882.02999\pm0.00138$ & 1\\
HAT-P-19b  &    FLWO 1.2 m   & $5111.57694\pm0.00102$ & 2$^{\textit{a}}$\\
HAT-P-19b  &    FLWO 1.2 m   & $5135.63142\pm0.00104$ & 2$^{\textit{a}}$\\
HAT-P-19b  &    FLWO 1.2 m   & $5163.69374\pm0.00111$ & 2$^{\textit{a}}$\\
HAT-P-19b  &    FLWO 1.2 m   & $5167.70166\pm0.00092$ & 2$^{\textit{a}}$\\
HAT-P-19b  &    Jena 0.6 m   & $5889.28345\pm0.00049$ & 3\\
HAT-P-19b  &    Jena 0.6 m   & $5905.31810\pm0.00044$ & 3\\
HAT-P-19b  &    CA-DLR 1.2 m & $5913.33571\pm0.00034$ & 3\\
HAT-P-19b  &    Jena 0.6 m   & $6935.57559\pm0.00055$ & 3\\
HAT-P-19b  &    GTC          & $5937.38839\pm0.00011$ & 4\\
HAT-P-19b  &    Toru\'{n} 0.6 m & $7300.37489\pm0.00040$ & 5\\
HAT-P-19b  &    Toru\'{n} 0.6 m & $7304.38284\pm0.00039$ & 5\\
\hline
HAT-P-51b  &    TCS          & $8408.58379\pm0.00019$ & 1\\
HAT-P-51b  &    TCS          & $8446.55780\pm0.00040$ & 1\\
HAT-P-51b  &    TCS          & $8699.62675\pm0.00021$ & 1\\
HAT-P-51b  &    TESS         & $8767.11668\pm0.00179$ & 1\\
HAT-P-51b  &    TESS         & $8771.32811\pm0.00186$ & 1\\
HAT-P-51b  &    TESS         & $8775.54648\pm0.00372$ & 1\\
HAT-P-51b  &    TESS         & $8779.76896\pm0.00179$ & 1\\
HAT-P-51b  &    TESS         & $8783.98563\pm0.00212$ & 1\\
HAT-P-51b  &    TESS         & $8788.20607\pm0.00247$ & 1\\
HAT-P-51b  &    TESS         & $9855.36415\pm0.00161$ & 1\\
HAT-P-51b  &    TESS         & $9859.58096\pm0.00155$ & 1\\
HAT-P-51b  &    TESS         & $9868.01820\pm0.00144$ & 1\\
HAT-P-51b  &    TESS         & $9872.23663\pm0.00162$ & 1\\
HAT-P-51b  &    TESS         & $9876.45429\pm0.00153$ & 1\\
HAT-P-51b  &    TESS         & $9880.67726\pm0.00157$ & 1\\
HAT-P-51b  &    FLWO 1.2 m   & $5856.68101\pm0.00071$ & 6$^{\textit{a}}$\\
HAT-P-51b  &    FLWO 1.2 m   & $5932.60419\pm0.00105$ & 6$^{\textit{a}}$\\
HAT-P-51b  &    FLWO 1.2 m   & $6206.77652\pm0.00098$ & 6$^{\textit{a}}$\\
HAT-P-51b  &    FLWO 1.2 m   & $6227.86597\pm0.00084$ & 6$^{\textit{a}}$\\
HAT-P-51b  &    FLWO 1.2 m   & $6244.74004\pm0.00103$ & 6$^{\textit{a}}$\\
\hline
HAT-P-55b  &    TCS          & $8254.55943\pm0.00046$ & 1\\
HAT-P-55b  &    TCS          & $8652.51935\pm0.00024$ & 1\\
HAT-P-55b  &    TCS          & $8713.46849\pm0.00025$ & 1\\
HAT-P-55b  &    TCS          & $9387.48890\pm0.00466$ & 1\\
\hline
\end{tabular}
}
\end{table}

\setcounter{table}{2}

\begin{table}
\renewcommand\arraystretch{1.5}
\centering
\caption{(Continued.)}
\label{tab: mid-transit time}
\scalebox{0.8}{
\begin{tabular}{cccc}
\hline\hline
Planet     & Telescope       & $T_{\rm mid} - 2450000 [\rm BJD_{TDB}]$ & Ref.	\\
\hline
HAT-P-55b  &    TESS         & $8985.94888\pm0.00233$ & 1\\
HAT-P-55b  &    TESS         & $8989.53122\pm0.00165$ & 1\\
HAT-P-55b  &    TESS         & $8993.11849\pm0.00172$ & 1\\
HAT-P-55b  &    TESS         & $9000.28713\pm0.00171$ & 1\\
HAT-P-55b  &    TESS         & $9003.87156\pm0.00181$ & 1\\
HAT-P-55b  &    TESS         & $9007.45673\pm0.00153$ & 1\\
HAT-P-55b  &    TESS         & $9011.04274\pm0.00144$ & 1\\
HAT-P-55b  &    TESS         & $9014.62923\pm0.00167$ & 1\\
HAT-P-55b  &    TESS         & $9018.21567\pm0.00189$ & 1\\
HAT-P-55b  &    TESS         & $9021.79611\pm0.00175$ & 1\\
HAT-P-55b  &    TESS         & $9025.38530\pm0.00158$ & 1\\
HAT-P-55b  &    TESS         & $9028.96794\pm0.00160$ & 1\\
HAT-P-55b  &    TESS         & $9032.55352\pm0.00160$ & 1\\
HAT-P-55b  &    TESS         & $9720.91828\pm0.00319$ & 1\\
HAT-P-55b  &    TESS         & $9724.50368\pm0.00194$ & 1\\
HAT-P-55b  &    TESS         & $9728.08945\pm0.00154$ & 1\\
HAT-P-55b  &    TESS         & $9731.67384\pm0.00156$ & 1\\
HAT-P-55b  &    TESS         & $9738.84360\pm0.00180$ & 1\\
HAT-P-55b  &    TESS         & $9742.42780\pm0.00115$ & 1\\
HAT-P-55b  &    TESS         & $9746.01622\pm0.00156$ & 1\\
HAT-P-55b  &    TESS         & $9753.18433\pm0.00150$ & 1\\
HAT-P-55b  &    TESS         & $9760.35539\pm0.00168$ & 1\\
HAT-P-55b  &    TESS         & $9763.94164\pm0.00175$ & 1\\
HAT-P-55b  &    TESS         & $9767.52606\pm0.00155$ & 1\\
HAT-P-55b  &    FLWO 1.2 m   & $6730.83546\pm0.00027$ & 7\\
HAT-P-55b  &    Haleadkala   & $8985.9487\pm0.0015$   & 8\\
HAT-P-55b  &    Haleadkala   & $9305.0323\pm0.0018$   & 8\\
\hline
HAT-P-65b  &    TCS          & $8329.54876\pm0.00038$ & 1\\
HAT-P-65b  &    TCS          & $8712.54036\pm0.00645$ & 1\\
HAT-P-65b  &    TESS         & $9799.01953\pm0.00174$ & 1\\
HAT-P-65b  &    TESS         & $9801.62787\pm0.00181$ & 1\\
HAT-P-65b  &    TESS         & $9804.23259\pm0.00192$ & 1\\
HAT-P-65b  &    TESS         & $9812.05096\pm0.00164$ & 1\\
HAT-P-65b  &    TESS         & $9814.65134\pm0.00191$ & 1\\
HAT-P-65b  &    TESS         & $9817.25814\pm0.00170$ & 1\\
HAT-P-65b  &    TESS         & $9819.86376\pm0.00209$ & 1\\
HAT-P-65b  &    FLWO 1.2 m   & $5739.73328\pm0.00083$ & 9$^{\textit{a}}$\\
HAT-P-65b  &    FLWO 1.2 m   & $5757.96850\pm0.00212$ & 9$^{\textit{a}}$\\
HAT-P-65b  &    FLWO 1.2 m   & $5825.71212\pm0.00205$ & 9$^{\textit{a}}$\\
HAT-P-65b  &    FLWO 1.2 m   & $6552.63474\pm0.00253$ & 9$^{\textit{a}}$\\
HAT-P-65b  &    FLWO 1.2 m   & $6565.66238\pm0.00079$ & 9$^{\textit{a}}$\\
HAT-P-65b  &    FLWO 1.2 m   & $6570.86994\pm0.00077$ & 9$^{\textit{a}}$\\
HAT-P-65b  &    GTC          & $8329.54829\pm0.00041$ & 10$^{\textit{a}}$\\
HAT-P-65b  &    GTC          & $9069.49471\pm0.00048$ & 10$^{\textit{a}}$\\
\hline
\end{tabular}
}
\begin{flushleft}
    \noindent{\footnotesize{Notes.$^{\textit{a}}$These mid-transit times have been recalculated using our light-curve analysis method.}}\\
    \noindent{\footnotesize{References.(1) This work. (2) \citet{2011ApJ...726...52H}. (3) \citet{2015MNRAS.451.4060S}. (4) \citet{2015AA...580A..60M}. (5) \citet{2018IBVS.6243....1M}. (6) \citet{2015AJ....150..168H}. (7) \citet{2015PASP..127..851J}. (8) \citet{2021MNRAS.504.5671E}. (9) \citet{2016AJ....152..182H}. (10) \citet{2021ApJ...913L..16C}.}}
\end{flushleft}
\end{table}

\begin{figure*}
\centering
\includegraphics[width=0.95\textwidth]{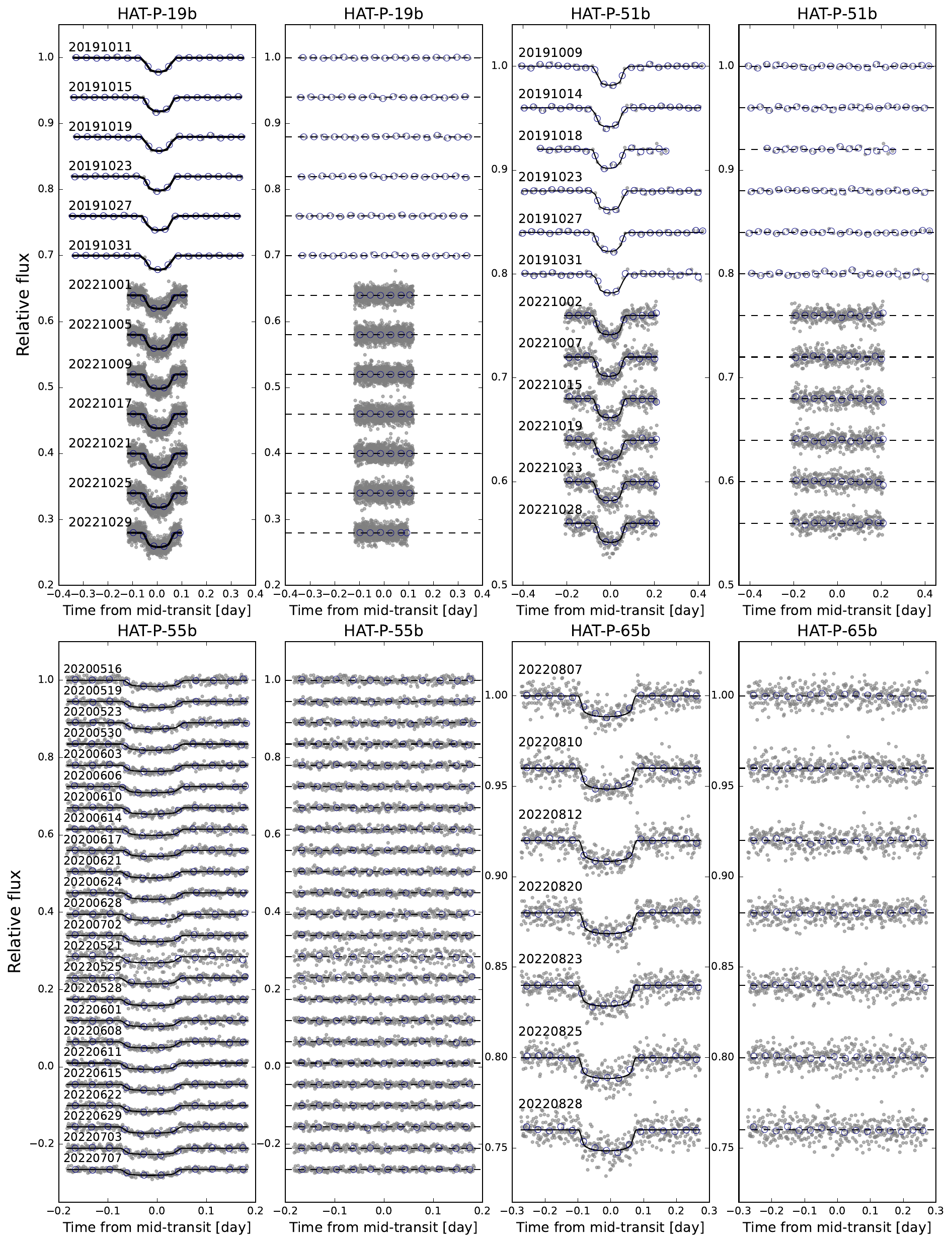}
\caption{TESS transit light curves of HAT-P-19b, HAT-P-51b, HAT-P-55b, and HAT-P-65b. The first and third panels show the light curves after removal of systematics and the solid lines show the best-fit model, the second and fourth panels show the best-fit residual, and the navy circles show the 60-min binned points.}
\label{fig: TESS light curves}
\end{figure*}


\bsp	
\label{lastpage}
\end{document}